\documentclass[twocolumn]{aastex631}

\usepackage{mathtools}
\begin{document}

\title{A Hidden Pulse: Uncovering a New Timing Signal in Cygnus X-1 with AstroSat}

\author[0000-0001-7590-5099]{Sandeep K. Rout}
\affiliation{New York University Abu Dhabi, PO Box 129188, Abu Dhabi, UAE}
\affiliation{Center for Astrophysics and Space Science (CASS), New York University Abu Dhabi, PO Box 129188, Abu Dhabi, UAE}

\author[0000-0001-9072-4069]{Federico Garc\'ia}
\affiliation{Instituto Argentino de Radioastronom\'a (CCT La Plata, CONICET; CICPBA; UNLP), C.C.5, 1894 Villa Elisa, Buenos Aires, Argentina}

\author[0000-0003-2187-2708]{Mariano M\'endez}
\affiliation{Kapteyn Astronomical Institute, University of Groningen, PO Box 800, NL-9700 AV Groningen, the Netherlands}

\author[0000-0001-5298-8303]{Abhay Kumar}
\affiliation{INAF Istituto di Astrofisica e Planetologia Spaziali, Via del Fosso del Cavaliere 100, 00133 Roma, Italy}

\author[0000-0002-2050-0913]{Santosh Vadawale}
\affiliation{Physical Research Laboratory, Navrangpura, Ahmedabad, Gujarat 380009, India
}

\author[0000-0002-3500-631X]{David M. Russell}
\affiliation{New York University Abu Dhabi, PO Box 129188, Abu Dhabi, UAE}
\affiliation{Center for Astrophysics and Space Science (CASS), New York University Abu Dhabi, PO Box 129188, Abu Dhabi, UAE}

\author[0009-0000-9011-7365]{Pei Jin}
\affiliation{Kapteyn Astronomical Institute, University of Groningen, PO Box 800, NL-9700 AV Groningen, the Netherlands}

\correspondingauthor{Sandeep K. Rout}
\email{sandeep.rout@nyu.edu}

\begin{abstract}

The study of fast variability properties in X-ray binaries advances our understanding of the physical processes and geometric properties of the accretion flow around the compact object. In this work, we study the evolution of the timing properties of Cygnus X-1 with AstroSat/LAXPC, during the transition of the source from the hard to soft state in 2017. We use a novel frequency-segmented technique to fit simultaneously the cross spectra and parts of the power spectra and coherence function with a multi-Lorentzian model and predict the phase-lags and the complementary parts of the power spectra and coherence function. We study the evolution of the frequency and power of the main variability components that are present throughout all the states. In particular, we identify previously undetected variability components, one of which manifests as a narrow dip in the coherence function and a broad drop in the phase-lag spectrum at the same frequency. This dip in coherence, which we detected for the first time in Cygnus X-1 at energies above 3 keV, appears in a state in which the source shows high-amplitude radio variability and significant hard X-ray polarization. While the contribution of the compact jet in X-rays is debated in the literature, this study provides a new avenue for investigating jet properties as well as the geometry of the Comptonizing medium.

\end{abstract}

\section{Introduction} \label{sec:intro}

Variability on timescales ranging from milliseconds to years reveals valuable information about X-ray binaries \citep[XRBs; see, e.g.,][for reviews]{mcclintock06, done07, kalemci22}. While long-timescale variability provides insight into the orbit of the system, the properties of the donor star, and the accretion-disk instabilities \citep[e.g.,][]{kotze12}, rapid variability is crucial for constraining the physical and geometric properties of matter in the vicinity of the compact object \citep[see, e.g.,][for reviews]{ingram19b, mendez21}. Fast variability in astrophysical sources is typically characterized by various features in the power spectra (PS), such as quasi-periodic oscillations (QPOs) and broadband noise \citep[BBN; e.g.,][]{vanderklis89a}. The PS provides key insights into the source's variability by capturing fluctuations in light curves across a range of frequencies. QPOs and BBN can vary in both frequency and amplitude depending on the source's physical properties and spectral state, making them valuable diagnostic tools for understanding the underlying mechanisms of accretion and emission. Spectral states, such as the hard and soft states, refer to distinct patterns in the time-averaged energy spectrum of the source, typically characterized by differences in luminosity and spectral hardness \citep[][]{vanderklis94}. In addition, the PS exhibits different characteristics in each state: for example, low-frequency QPOs and flatter BBN are common in the hard state, while high-frequency QPOs and steeper BBN are observed in the soft state \citep[e.g.,][]{belloni02}. Extensive observational studies over the last few decades have refined these connections, providing a thorough understanding of the interplay between variability, spectral states, and accretion flow geometry \citep[e.g.,][]{homan01, belloni02, casella04, grinberg14, rout21a, mendez22, rout25}.

Apart from the PS, the time lags and coherence function - derived from the cross spectrum (CS) of correlated light curves in two energy bands - also provide independent information on the source. The time lags represent the relative delay between the arrival times of photons in the two energy bands. The evolution of this lag with Fourier frequency and energy serves as a useful diagnostic for determining both the emission mechanism and the geometry of the emitting media \citep[e.g.,][]{arevalo06, ingram11, kara19, kylafis20, karpouzas20}. The coherence function, on the other hand, measures the degree of linear correlation between signals in the two energy bands \citep{vaughan97, bendat10}, and helps to identify the range of frequencies that are strongly correlated in the energy bands \citep{nowak99a}. When the relation between the emission in the two energy bands is non linear, or multiple independent processes are at play, the coherence falls below unity \citep{bendat10, vaughan97, nowak99a}. Similar to the PS, both the lag spectrum and coherence function are known to change as functions of the spectral states. Recent studies have detected sharp drops in the coherence in the $\sim1-6$ Hz frequency range in MAXI J1820$+$070 and Cygnus X-1 using NICER data \citep[e.g.,][]{mendez24, konig24}. These drops were interpreted as the result of the presence of a variability component that is significant in the CS - especially in the imaginary part - but not in the PS \citep{mendez24}. Because the drop in coherence appeared only when one of the energy bands was below 2 keV, it was associated with the variability of the accretion disk \citep{konig24, fogantini25}.

Cygnus X-1 (hereafter Cyg X-1) hosts a $21.2\pm2.2~ M_\odot$ black hole (BH) that accretes matter from the stellar wind of a supergiant companion \citep{walborn73, millerjones21}. Being a bright and persistent source, it has become the subject of numerous studies, serving as a laboratory to test various ideas \citep[e.g.,][]{dove97, arevalo06, dexter24}. In fact, frequent observations of Cyg X-1 with earlier missions led to the discovery of spectral state transitions \citep{tananbaum72}. Since then, a vast amount of literature has focused on studying the evolution of spectral, timing, and polarization properties as functions of the spectral states \citep[e.g.,][]{cui97b, dove98, axelsson05, wilms06, jourdain12, zhoum22, krawczynski22, chattopadhyay24, konig24}. However, it is important to note that the spectral states in Cyg X-1 are not strictly equivalent to the states commonly observed during the outbursts of transient BH low-mass X-ray binaries \citep[LMXBs;][]{mcclintock06, done07, belloni10, belloni16}. In particular, unlike the high soft state (HSS) in other LMXBs, the soft state in Cyg X-1 shows high levels of variability ($\gtrsim 20\%$) and a strong hard X-ray Comptonization component \citep[e.g.,][]{grinberg14, wilms06}. Also, despite being a bright source with a high power-law flux extending to very high energies, Cyg X-1 has so far never shown QPOs in any phase of accretion. The absence of QPOs, along with the fact that the variability amplitude is always high, makes it difficult to subclassify the intermediate state into the established hard-intermediate state (HIMS) and soft-intermediate state \citep[SIMS;][]{belloni05, belloni16}. Therefore, one must be cautious when extrapolating the results pertaining to spectral states in Cyg X-1 to the general population of BH LMXBs. Moreover, with a long-term study with INTEGRAL \citet{lubinski20} showed that the states in Cyg X-1 could be subdivided into six different accretion modes.  

In this work, we study the evolution of the power and cross-spectral properties of Cyg X-1 using data from Large Area X-ray Proportional Counter (LAXPC) onboard AstroSat, as the source transitions from the hard to the soft state. In Section \ref{sec:analysis}, we describe the data reduction process and analysis methods used (further details are provided in the Appendix). The main findings of this work are presented in Section \ref{sec:result}. Finally, in Section \ref{sec:discussion}, we interpret the results in the context of different physical models and propose possible scenarios. 

\begin{table*}[]
    \centering
    \begin{tabular}{c|c|c||c|c|c}
        Observation ID & Date/ MJD & Exp. (s) & Observation ID & Date/ MJD & Exp. (s)\\
        \hline
         G06\_028T01\_900000\textbf{0890} & 2016 Dec 16/ 57738.6 & 6702 & G07\_027T01\_900000\textbf{1304} & 2017 Jun 15/ 57919.5 & 2064 \\ 
         G06\_028T01\_900000\textbf{1094} & 2017 Mar 20/ 57832.8 & 10768 & G07\_027T01\_900000\textbf{1358} & 2017 Jul 05/ 57939.5 & 10512 \\
         G06\_028T01\_900000\textbf{1122} & 2017 Mar 31/ 57843.7 & 14032 & G07\_027T01\_900000\textbf{1470} & 2017 Aug 17/ 57982.7 & 9952 \\
         G07\_027T01\_900000\textbf{1180} & 2017 Apr 17/ 57860.8 & 11008 & G07\_027T01\_900000\textbf{1540} & 2017 Sep 17/ 58013.4 & 11424 \\
         G07\_027T01\_900000\textbf{1210} & 2017 May 08/ 57881.4 & 10992 & G08\_030T01\_900000\textbf{1592} & 2017 Oct 08/ 58034.0 & 4784 \\
    \end{tabular}
    \caption{List of observations used in this work, along with the dates and exposure times (in seconds) of LAXPC data used for the timing analysis.}
    \label{tab:obs}
\end{table*}

\section{Analysis} \label{sec:analysis}

PS in XRBs are typically modeled by a linear combination of Lorentzian functions \citep{miyamoto91}. The usual interpretation \citep[e.g.,][]{belloni02} is that these Lorentzians represent different variability components in the system, corresponding to distinct processes or properties of the emitting media. \citet{mendez24} showed that, under the assumption that the individual Lorentzians are coherent in different energy bands but incoherent with one another, the Lorentzians that fit the PS in two energy bands also fit the CS (both the real and imaginary parts) from the same energy bands (see Appendix \ref{sec:method} for more details on the method). The strength of this technique is that the fitted Lorentzians can be used to predict the phase-lags and coherence spectra (Appendix \ref{sec:method}). Since the six spectra - the PS in two energy bands, the real and imaginary parts of the CS, the phase lags, and the coherence function - are not all independent, we fit only four spectra and predict the other two. 

In some cases, it is more useful to fit different parts (i.e. in different frequency ranges) of the six spectra, ensuring that the total number of degrees of freedom remains the same as if only four spectra were fitted. From Equation \ref{eq:coh}, it is straightforward to see that instead of predicting the coherence and phase lags, one can fit the coherence along with one PS (e.g., in the subject band) and the real and imaginary parts of the CS to predict the second PS (e.g., in the reference band) and the phase lags. This is because when power is high the uncertainty in the coherence function, given by $\sigma_{\gamma^2} \simeq \sqrt{2}(1-\gamma^2_{xy})|\gamma_{xy}|/\sqrt{N}$ \citep{bendat10}, decreases rapidly as the value of coherence approaches unity. This means that some variability components can have a larger signal-to-noise ratio (SNR) in the coherence than in the PS and CS. However, at higher frequencies (especially $>10$ Hz), the uncertainty in the coherence increases rapidly as the power decreases and the noise becomes significant \citep[see][]{vaughan97}, while at those frequencies the PS still has a significant power. Therefore, to maximize the detection of all components, we employ for the first time a frequency-segmented fitting approach. We fit the subject-band PS in the $0.3-100$ Hz range and the coherence-frequency spectrum in the $0.002-0.3$ Hz range, along with the reference-band PS and the CS between the two bands in the full frequency range ($0.002-100$ Hz). We then predict the subject-band PS in the remaining $0.002-0.3$ Hz range and the coherence function in the $0.3-100$ Hz range, along with the phase-lag spectrum over the full frequency range.

LAXPC data of sources with high count rate are affected by deadtime, which depends on the type of events \citep{yadav16b}. Therefore, the traditional method of correcting Poisson noise by subtracting the average power from a high frequency range does not work for higher-order statistics such as coherence \citep[e.g.,][]{vaughan97}. In order to circumvent the effects of deadtime, we use, for the first time, the cross spectra between two LAXPC units (10 and 20) for all the Fourier products. This method was proposed by \citet{bachetti15} to address the effects of dead time in the PS obtained from NuSTAR. Here we not only extract the PS, but also the CS, phase lags and coherence from the cross spectra of the two detectors (see Appendix \ref{sec:deadtime} for detailed expressions of the Fourier products). This is crucial, as deadtime introduces correlations between channels, which artificially reduces the coherence at high Fourier frequencies. 

AstroSat made several observations of Cyg X-1 during the transition from the hard to soft state throughout 2017. In this work, we use a subset of these observations made between 16 December 2016 (MJD 57738) and 08 October 2017 (MJD 58034). The list of observations is presented in Table \ref{tab:obs}. These data were chosen to study the evolution of the timing properties in almost all levels of flux and hardness of the source. We also carried out a joint spectral fit with all the three X-ray instruments onboard AstroSat for one observation (1210) to have an estimate of the flux fractions of the individual components in the reference and subject bands (Appendix \ref{sec:spec}). The reduction of LAXPC data \citep{yadav16a} was performed using the Format-A\footnote{\url{http://astrosat-ssc.iucaa.in/laxpcData}} pipeline following the standard procedures. The level 2 event files were used to generate the Fourier products using \texttt{GHATS}\footnote{\url{http://astrosat-ssc.iucaa.in/uploads/ghats_home.html}}. We computed all the Fourier products in the $3-5$ keV and $6-40$ keV bands; for the CS we took the former as the reference and the latter as the subject band. We rebinned the LAXPC lightcurve, having a resolution of $10~\mu s$, by a factor of 50 to achieve a Nyquist frequency of 1000 Hz. Fast Fourier transforms (FFTs) were performed on data segments of length $\approx 524.29$ s giving a frequency resolution of $\sim 0.0019$ Hz, except for observation 1304 for which segments of $\approx 262.14$ s were used. To obtain a good SNR we ensured that FFTs from at least 10 segments were averaged in all the observations. 

\begin{figure*}
    \centering
    \includegraphics[scale=0.55]{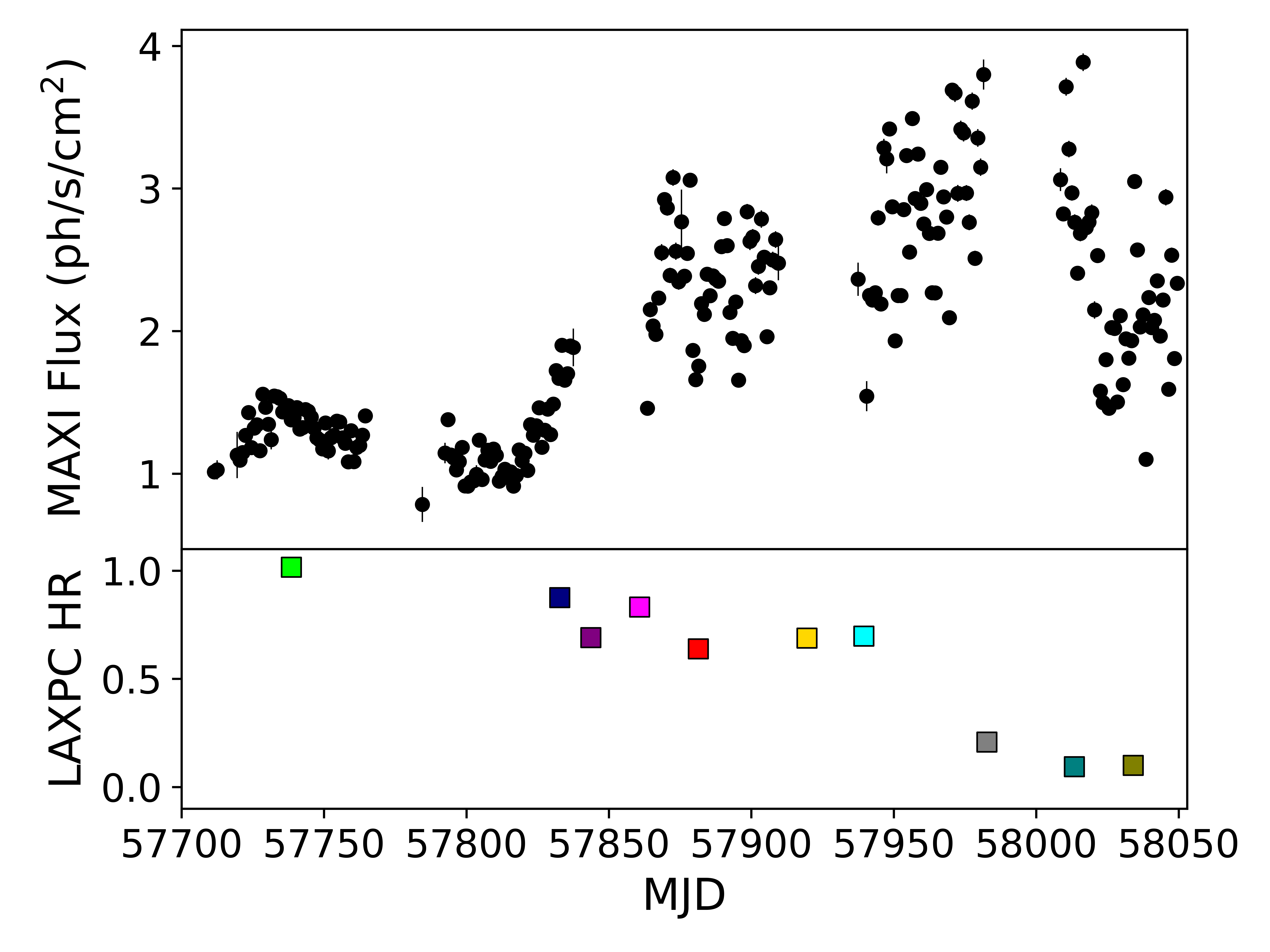}
    \includegraphics[scale=0.55]{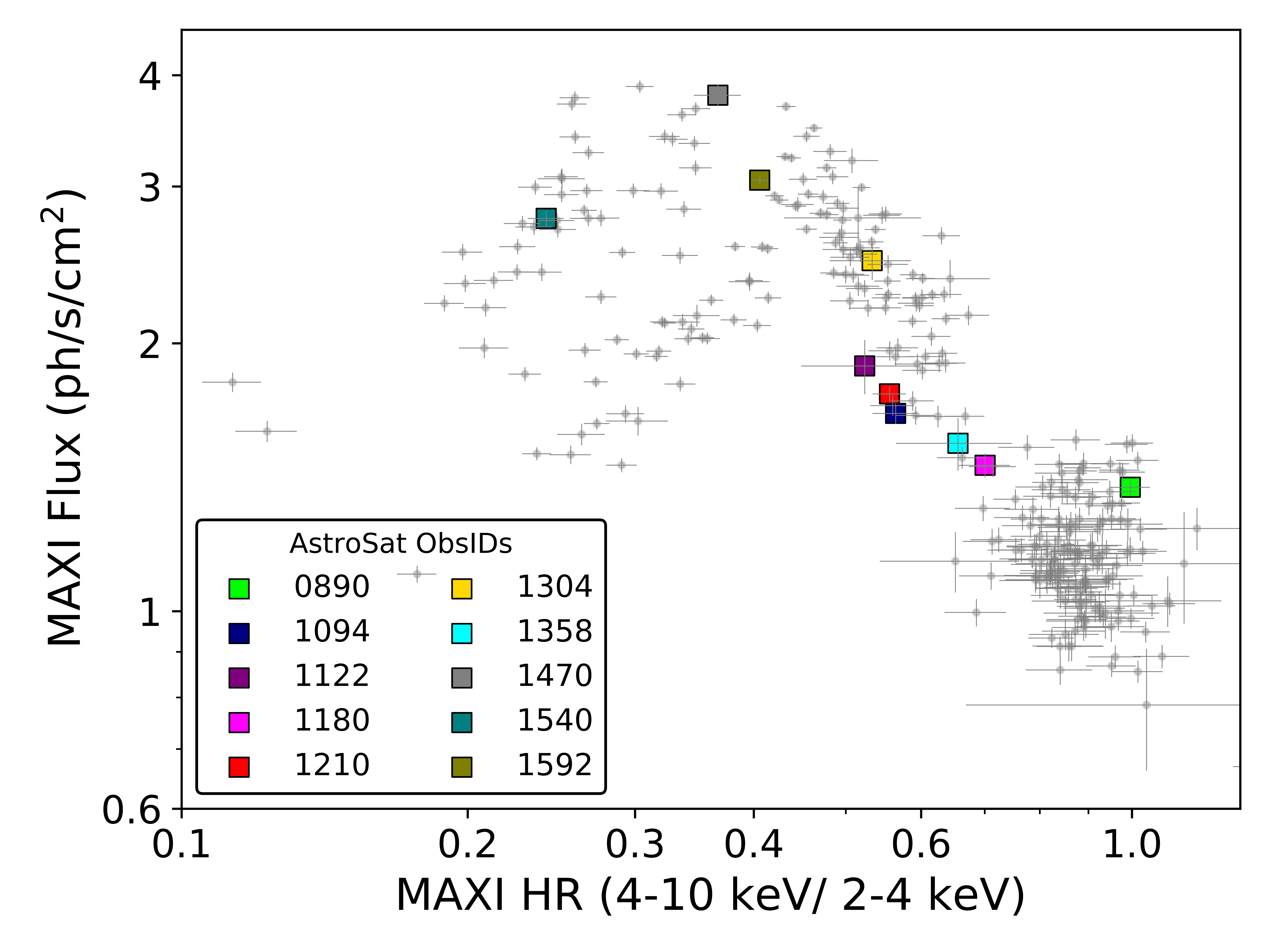}
    \caption{The top left panel shows the $2-20$ keV lightcurve of Cyg X-1 with MAXI. The bottom left panel shows the hardness ratio (HR) curve of Cyg X-1 with LAXPC data, defined as the ratio of count rates in the $15-30$ keV and $3-5$ keV bands. The right plot shows the hardness-intensity diagram of Cyg X-1 derived from MAXI lightcurve. The MAXI points closest to the epochs of AstroSat observations are highlighted.}
    \label{fig:lchr}
\end{figure*}

We have also carried out hard X-ray spectroscopy with the Cadmium Zinc Telluride Imager \citep[CZTI;][]{vadawale16} of all the ten observations to determine the exact spectral states as defined by \citet{lubinski20}. The CZTI spectra were extracted with the \texttt{cztpipeline\_v3.0} pipeline. The pipeline corrects for the background emission in the spectra by using a mask-weighting technique. It also generates the response files used for fitting. Spectra from quadrant 0 of the detector in the $30-100$ keV range were used for spectral analysis.

\section{Results} \label{sec:result}

The $2-20$ keV light curve of Cyg X-1, using data from the Monitor of All-sky X-ray Image \citep[MAXI;][]{matsuoka09}, is shown in the top left panel of Figure \ref{fig:lchr}. The bottom left panel displays the average hardness ratio (HR), calculated using LAXPC ($15-30$ keV / $3-5$ keV), for the 10 observations analyzed in this work. The spectral transition begins in March 2017 (around MJD 57825), characterized by an increase in the flux and a decrease in the HR. This transition continues for approximately eight months, after which the source reaches the soft state in October 2017. Observations 0890 and 1540 exhibit the highest and lowest HRs, respectively. The long-term hardness-intensity diagram (HID) of Cyg X-1 is shown in the right panel of Figure \ref{fig:lchr}, with the points corresponding to the 10 AstroSat epochs highlighted. Our observations cover nearly the entire range of HRs attained by the source. Figure \ref{fig:lchr} also shows that the HR does not evolve monotonically with the flux for Cyg X-1.

\begin{figure*}
    \centering
    \includegraphics[scale=.72]{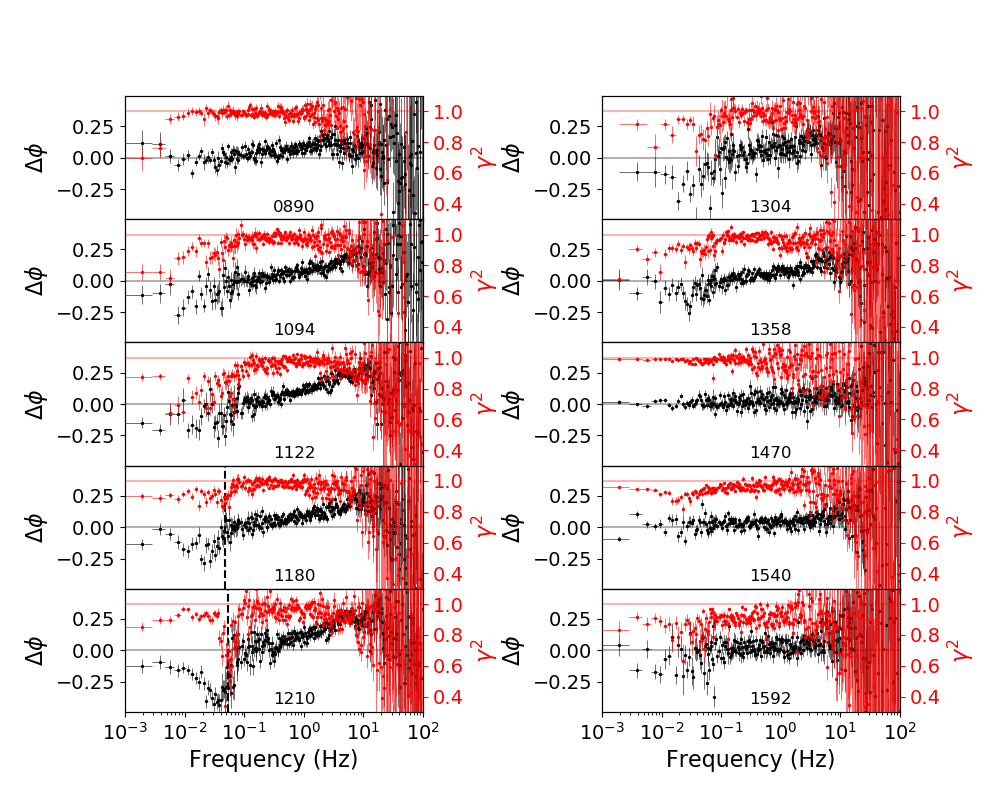}
    \caption{The phase lag frequency spectra (black) and coherence spectra (red) for all the ten observations of Cyg X-1 used in this work. The respective observation IDs are mentioned inside the panels. Observations 1180 and 1210 show the narrow drop in coherence, at frequencies marked by a dashed vertical line, accompanied by the fall in phase lag spectrum.}
    \label{fig:lagcoh}
\end{figure*}

In Figure \ref{fig:lagcoh}, we show the phase-lag frequency and coherence frequency spectra with $3-5$ keV as the reference band and $6-40$ keV as the subject band for all the observations. Although we show and fit the data in the $10^{-3}-10^2$ Hz range, both the coherence and phase lags are accurately measured only up to $\sim10$ Hz. Beyond this, the Poisson noise dominates the data. In observation 0890, the coherence remains close to unity in the $\sim 0.01-10$ Hz range and smoothly decreases to $\sim 0.6$ at lower frequencies. The phase lags remain near zero above 0.01 Hz and increase slightly below that. In subsequent observations, from 1094 to 1358, the coherence is non-unity up to 0.1 Hz, with almost unity coherence beyond that. These observations show strong hard lags at higher frequencies, peaking at $\sim 10$ Hz, and soft lags at frequencies $\lesssim 0.1$ Hz. The coherence in observation 1210 exhibits a sharp dip at $\sim 0.05$ Hz, dropping to $\sim 0.6$, accompanied by a fall in the phase lags to $\sim -0.4$ rad. Observation 1180 also shows a similar, though less pronounced, feature in both the coherence and phase lags. The last three observations (1470, 1540, and 1592) exhibit nearly unity coherence and zero phase lag throughout the entire frequency range.      

\subsection{State Classification}

\begin{figure}
    \centering
    \includegraphics[scale=.3]{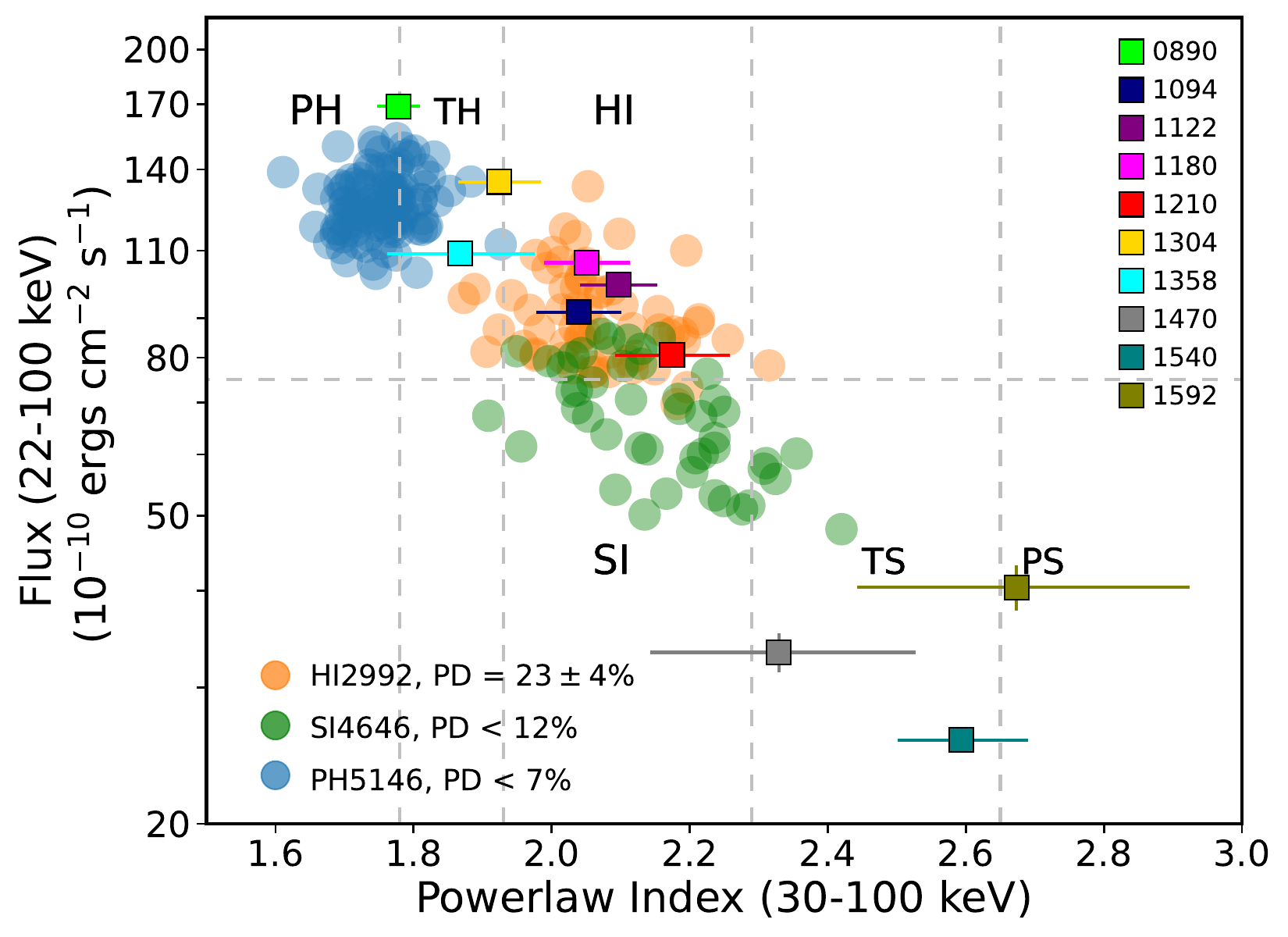}
    \caption{The squares show the $22-100$ keV flux plotted against the photon index ($30-100$ keV) obtained from fitting the CZTI spectra for all the 10 observations of Cyg X-1. The circles are from Figure 1 of \citet{chattopadhyay24} showing 3 observations with CZTI, with the legends displaying their respective observation IDs and polarization degrees (PD). The additional variability components, including the narrow dip at 0.05 Hz, appear only in the hard intermediate (HI) state, where the PD is $23\%$. In the other cases, \citet{chattopadhyay24} find only upper limits to PD that are significantly smaller than the PD value in the HI state.}
    \label{fig:czti}
\end{figure}

To determine the spectral states of the observations, as defined in \citet{lubinski20}, we carried out spectroscopy with CZTI \citep[see][]{chattopadhyay24}. The spectra of all the observations in the $30-100$ keV band were fitted by a single \texttt{powerlaw} component. The best-fitting values of the photon index and flux in the $22-100$ keV band are represented in Figure \ref{fig:czti}. The ten observations analyzed in this work broadly cover all the six states, or accretion modes. Using the state classification of \citet{lubinski20}, observation 0890 lies in the pure hard (PH) state, 1304 and 1358 lie in the transitional hard (TH) state, and 1094, 1122, 1180 and 1210 lie in the hard intermediate (HI) state. The last three observations - 1470, 1540, and 1592 - remain in the soft intermediate (SI), transitional soft (TS), and pure soft (PS) states, respectively, although with some uncertainty. 

\begin{figure*}
    \centering
    \includegraphics[scale=.49]{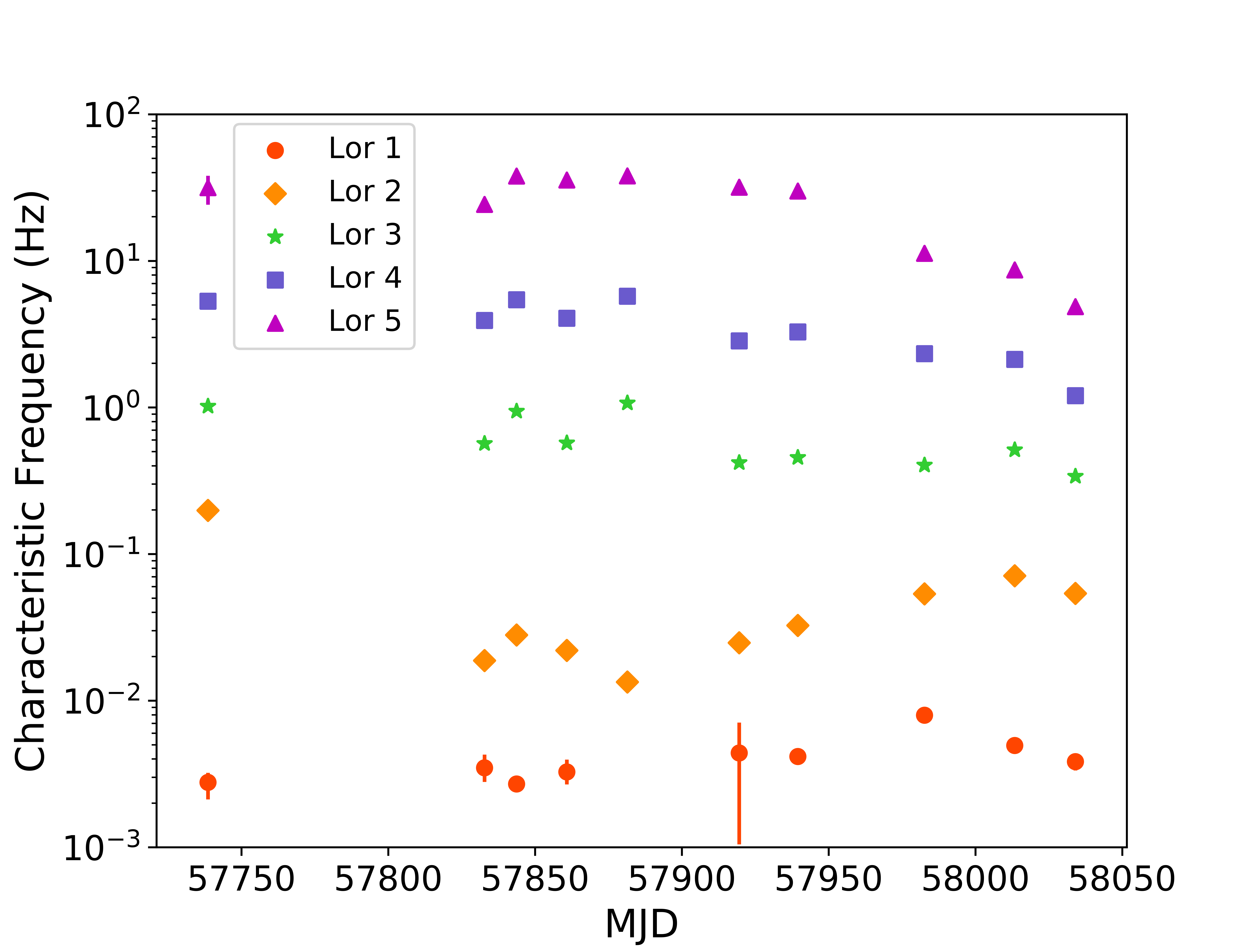}
    \includegraphics[scale=.49]{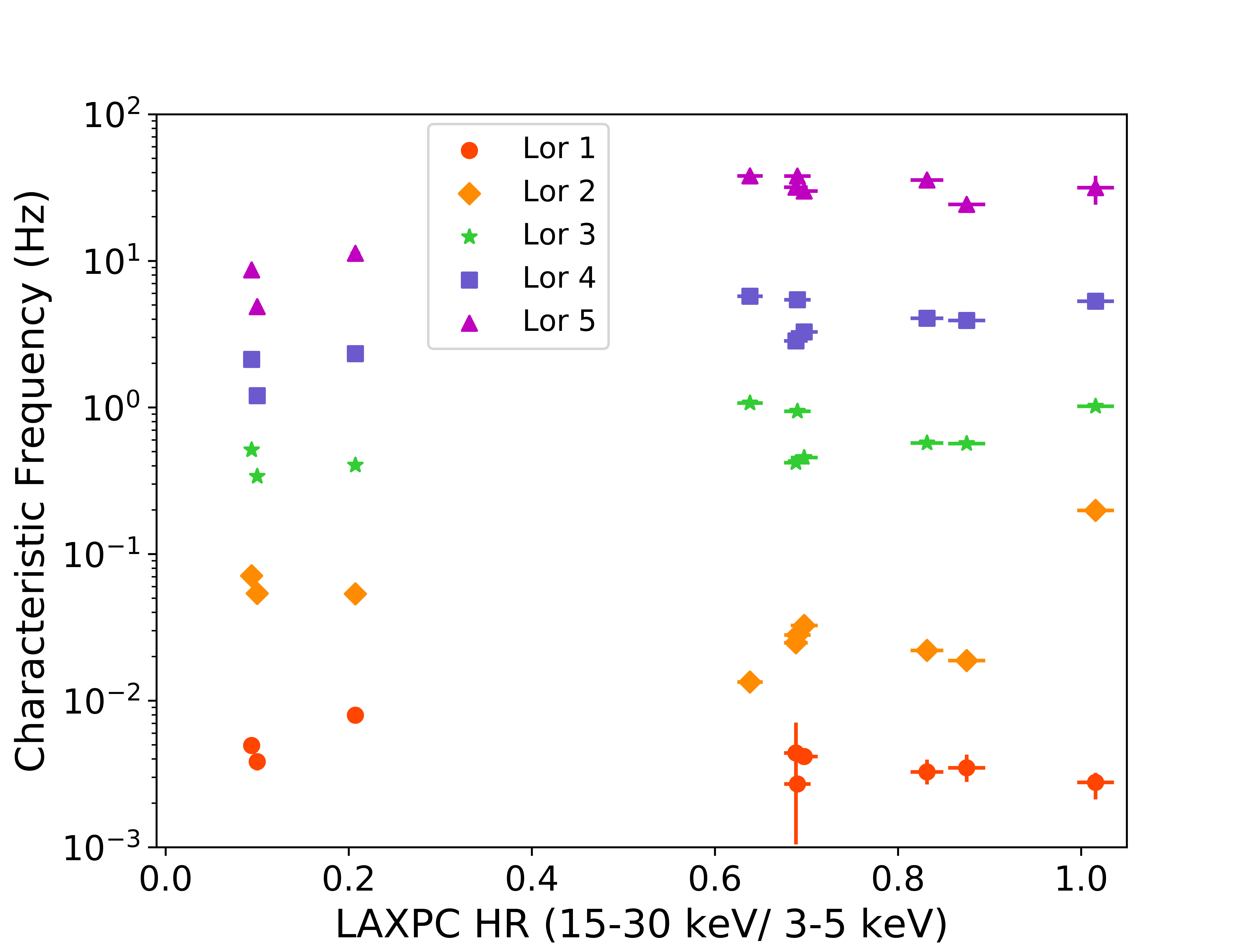}
    \caption{Left: The time evolution of the characteristic frequencies of the five main variability components in the ten observations of Cyg X-1. Right: The characteristic frequencies of the five variability components as a function of the HR, defined by the ratio of count rates in the $15-30$ keV and $3-5$ keV bands using LAXPC. The variability components are numbered in order of increasing frequency.}
    \label{fig:numax}
\end{figure*}

\subsection{Main variability components} \label{sec:mainlor}

As described in Appendix \ref{sec:method}, we performed simultaneous frequency-segmented fits with a multi-Lorentzian model to parts of the six spectra, i.e., the two PS, the real and imaginary parts of the CS, phase lags, and the coherence function, such that the total number of degrees of freedom is the same as in any four spectra. The PS in the reference and subject bands were fitted in, respectively, the $0.002-100$ Hz and $0.3-100$ Hz range as given by Equation \ref{eq:ps}. The real and imaginary parts of the CS were fitted in the $0.002-100$ Hz range using Equation \ref{eq:cs}, and the coherence function was fitted in the $0.002-0.3$ Hz range as defined in Equation \ref{eq:coh}. Using the same parameters of the fitted multi-Lorentzian model, the phase lags in the $0.002-100$ Hz range, the coherence function in the $0.3-100$ Hz range and the subject-band PS in the $0.002-0.3$ Hz range were compared with the predictions of Equations \ref{eq:lag}, \ref{eq:coh} and \ref{eq:ps}, respectively. From our simultaneous fits, we find that all the observations, including those in the soft states, require a minimum of five Lorentzians. Four observations, from 1094 to 1210, require some additional components that will be discussed in the next section.

The characteristic frequencies ($\nu_{max}=\sqrt{\nu_0^2+\left(\frac{\Delta}{2}\right)^2 }$, where $\nu_0$ is the centroid frequency and $\Delta$ is the full width at half maximum of the Lorentzians) of each of the main Lorentzians lie roughly in each dex between $10^{-3} - 10^2$ Hz. The time evolution of $\nu_{max}$ for the five main components is shown in Figure \ref{fig:numax}, where the components are numbered $1-5$ with increasing frequency. The two lowest-frequency components (Lor 1 and 2) show some variability in frequency. The three highest-frequency components (Lor 3, 4 and 5) remain mostly stable in the first five observations and shift toward lower frequencies in the rest of the observations. In the right panel of Figure \ref{fig:numax}, we show $\nu_{max}$ as a function of the HR evaluated as the ratio of the count rates in the $15-30$ keV band to the $3-5$ keV band. The three high-frequency Lorentzians (Lor 3, 4 and 5) show a slight increasing trend with hardness, while the low-frequency components (Lor 1 and 2) show a decreasing trend. 

\begin{figure*}
    \centering
    \includegraphics[scale=.55]{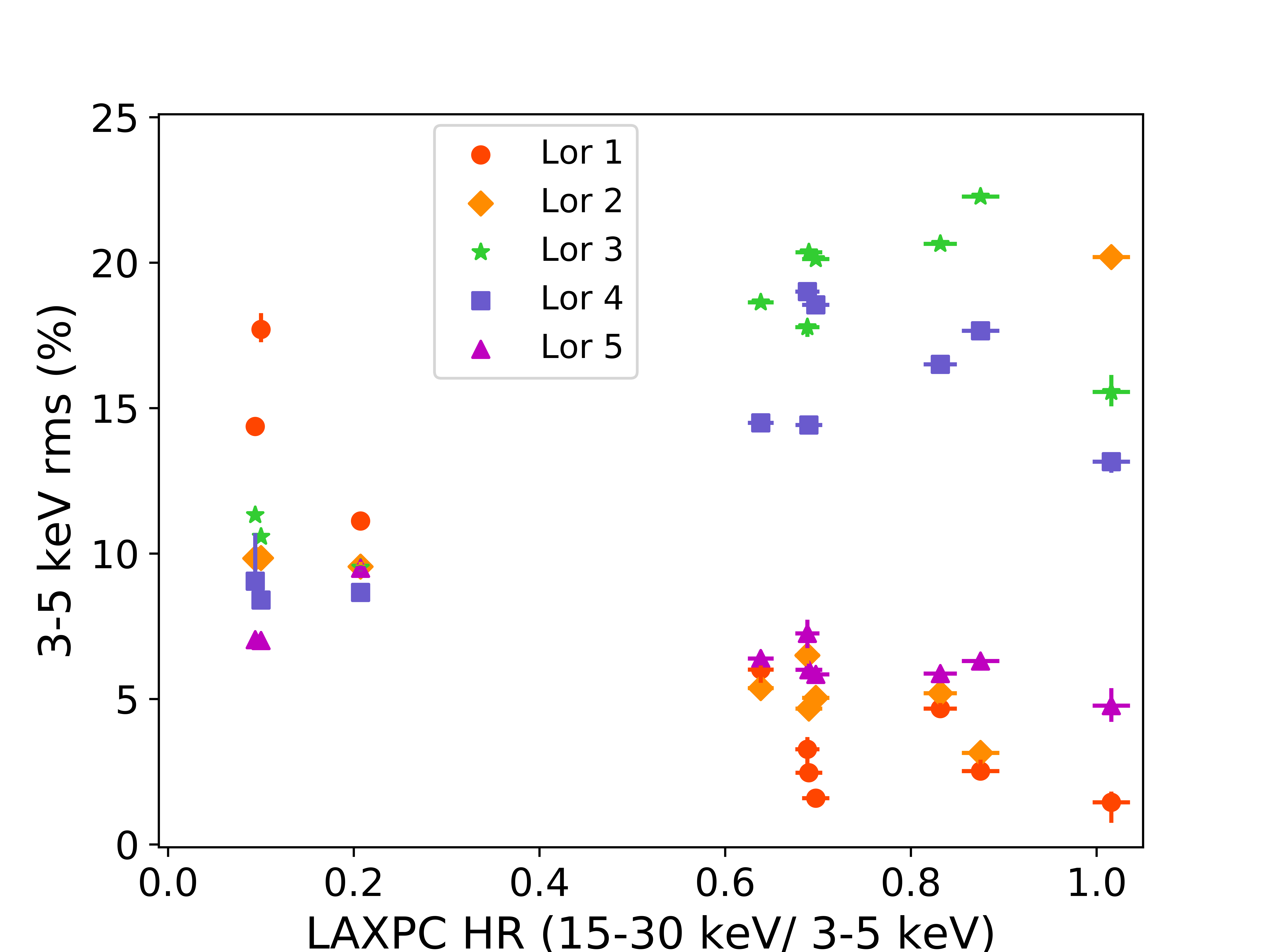}
    \includegraphics[scale=.55]{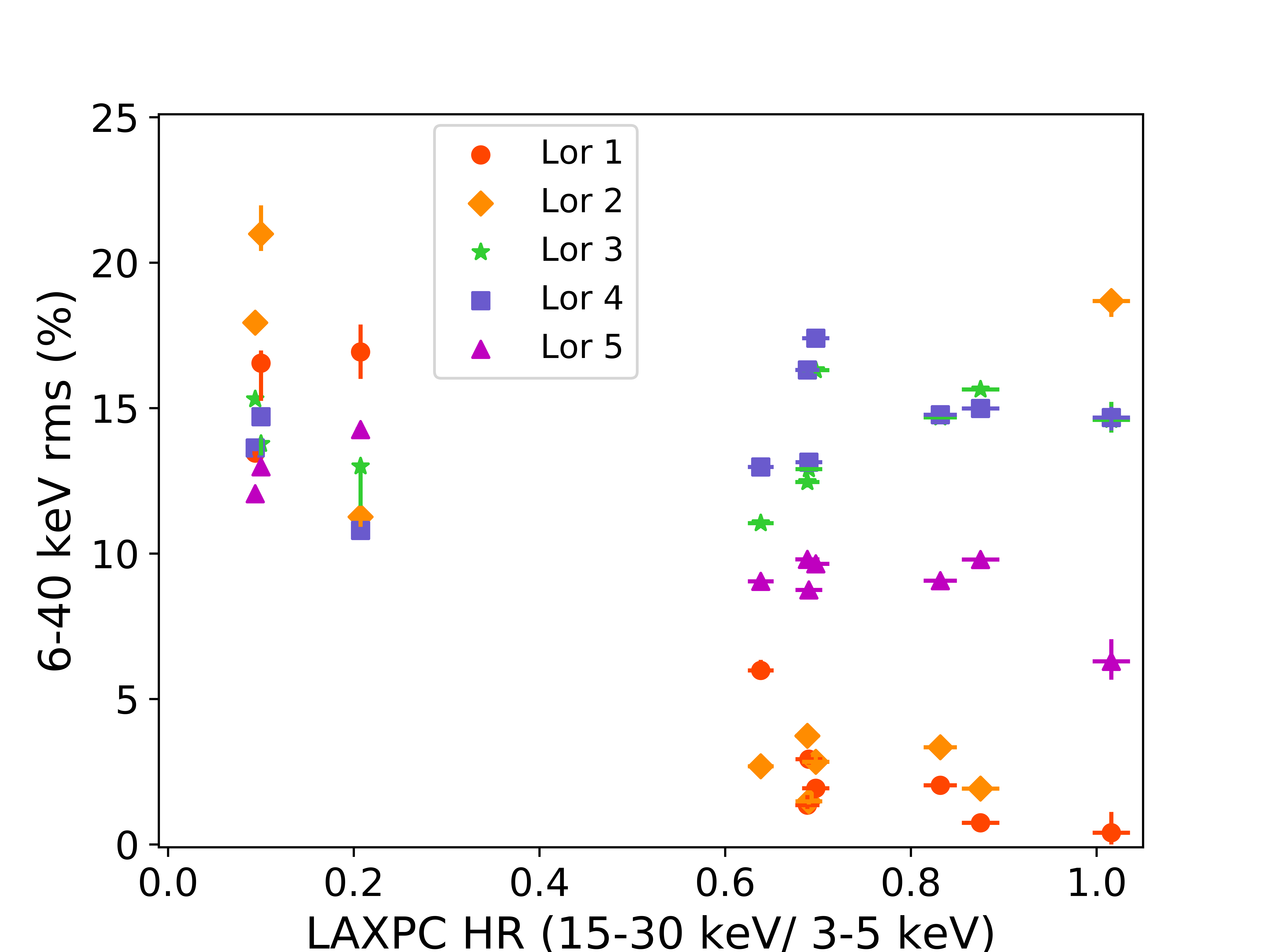}
    \caption{Fractional rms amplitude (in percentage) in the $3-5$ keV (Left) and $6-40$ keV (Right) bands as a function of HR for the five main Lorentzians in the ten observations of Cyg X-1.}
    \label{fig:rmsevo}
\end{figure*}

Figure \ref{fig:rmsevo} shows the evolution of the rms amplitude of the five Lorentzians in the $3-5$ keV and $6-40$ keV bands as a function of the HR. In both the energy bands, the rms of Lor 3 and Lor 4 increase with hardness, while for the other three components, the rms decrease with hardness. It is also apparent that different components dominate the total power in different observations and energy bands. When the HR is at its lowest (HR $<0.3$), the two lowest-frequency components dominate the total power (rms $\sim 10-20~\%$). As the hardness increases to intermediate values (HR $\sim 0.6-0.9$), the rms amplitude of the two lowest-frequency components (Lor 1 and Lor 2) decreases to $\lesssim5~\%$ in both the energy bands. In these intermediate states, Lor 3 and Lor 4 dominate the total power with rms $\sim15-20~\%$. In the observation with the highest HR (HR $>1.0$), Lor 2 ($\sim20~\%$) dominates the total power, followed by Lor 3 and Lor 4 ($\sim15~\%$) in both the energy bands. Notably, in the soft states (HR $<0.3$), the rms amplitude of all the five components remains $>10~\%$ in the $6-40$ keV band and $>6~\%$ in the $3-5$ keV band.

\subsection{Additional variability components} \label{sec:addlor}

Our frequency-segmented fitting framework, which simultaneously analyzes different Fourier spaces, has the potential to reveal previously unknown variability components (Section \ref{sec:analysis}). As mentioned in the previous section, the four observations - 1094, 1122, 1180 and 1210 - necessitate the inclusion of additional Lorentzians for fitting. These additional components play a crucial role in predicting the complex structures of the phase-lags and coherence spectra (Figure \ref{fig:lagcoh}). In all these four observations, there is a broad Lorentzian with $\nu_{max}$ lying between 0.03 Hz and 0.3 Hz. The rms amplitude of this component remains in the range of $\approx 4-6~\%$ in the $3-5$ keV band and $\approx 2-5~\%$ in the $6-40$ keV band. The F-statistic values for six Lorentzians, compared to the baseline model containing five Lorentzians, for observations 1094, 1122 and 1180 are $\sim4.4$, $\sim12.5$ and $\sim6.1$ respectively. The corresponding p-values are $\sim5.4\times10^{-4}$, $\sim5.7\times10^{-10}$ and $\sim1.3\times10^{-5}$, suggesting a significant improvement to the fits with this additional broad component. 

The coherence in observation 1180 shows a narrow dip at $\sim 0.05$ Hz at which the coherence drops down to $\sim 0.8$, and is also accompanied by a broad drop in the phase lags around the same frequency. This drop requires an extra narrow Lorentzian with $\nu_{max} = 0.047 \pm 0.002$ Hz\footnote{The errors here, and in the rest of the paper, correspond to $68~\%$ confidence range} and is highly significant ($5\sigma$) in the subject band ($6-40$ keV), but not in the reference band ($3-5$ keV; $1\sigma$). The rms of this component is $0.2\pm0.1~\%$ and $1.0\pm 0.1~\%$ in the reference and subject bands, respectively. The F-statistic and p-value for seven Lorentzians compared to the model with six components are $\sim3.2$ and $\sim6.2\times10^{-3}$, respectively. We note that the F-statistics and p-values reported here refer to improvements in the combined fits across all Fourier products.

\begin{figure*}
    \centering
    \includegraphics[scale=.6]{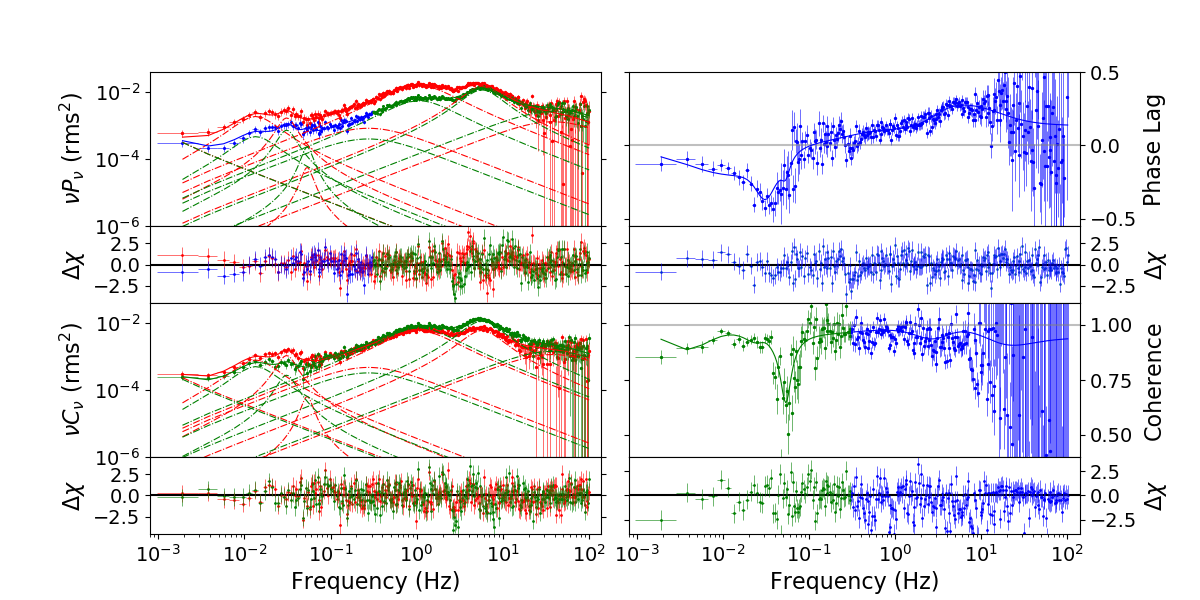}
    \caption{Data and the best-fitting multi-Lorentzian model, with residuals, for observation 1210 of Cyg X-1. The top left panels show the power spectra in the reference ($3-5$ keV; red) and subject ($6-40$ keV; blue and green) bands. The bottom left panels show the real (red) and imaginary (green) parts of the cross spectrum, both rotated by $45^\circ$. The dashed vertical lines in these two panels represent the characteristic frequency of the Lorentzian that fits the dip in the coherence. The top and bottom panels in the right-hand side show the phase-lag spectrum and the coherence function, respectively. The fitting is done to the data shown in red and green points, whereas the prediction is done to the data plotted in blue (see Appendix \ref{sec:method} for details). Solid lines represent the total model and dashed dotted lines represent individual components.}
    \label{fig:pclc1210}
\end{figure*}

In the next observation (1210), the drop in coherence is stronger, down to $\sim0.6$, and the phase lags reaches $\sim -0.4$ rad (Figure \ref{fig:pclc1210}). The Lorentzian that accounts for this drop has a best-fitting $\nu_{max}=0.053 \pm 0.001$ Hz and rms of $0.5\pm 0.2~\%$ and $1.01^{+0.06}_{-0.04}~\%$ for the two energy bands. Here also the component is significant ($\sim10\sigma$) only in the $6-40$ keV band. The model with this component results in a drastic improvement in the fit, with the F-statistic and the p-value being $\sim16.26$ and $\sim1.4\times10^{-15}$, respectively. The PS and CS in this observation show a double hump feature between $0.01-0.04$ Hz that requires another Lorentzian in addition to the usual Lor 2. This Lorentzian has a $\nu_{max} \sim 0.03$ Hz and rms of $3.5\pm0.2~\%$ and $2.3\pm0.2~\%$ in the two energy bands, with significances of $\sim8.7\sigma$ and $\sim5.7\sigma$, respectively. The F-statistic and p-value after the addition of this component are $\sim4.7$ and $\sim9.1\times10^{-4}$, respectively. Finally, we added the broad Lorentzian, which was detected in the previous three observations. This also significantly improved the fit with a F-statistic and p-value of $\sim5.6$ and $\sim8.2\times10^{-6}$, respectively, compared to the model with seven components. It is interesting to note that the four observations that require additional variability components, including the dip in the coherence, lie in the same state (HI).

\begin{figure}
    \centering
    \includegraphics[scale=.43]{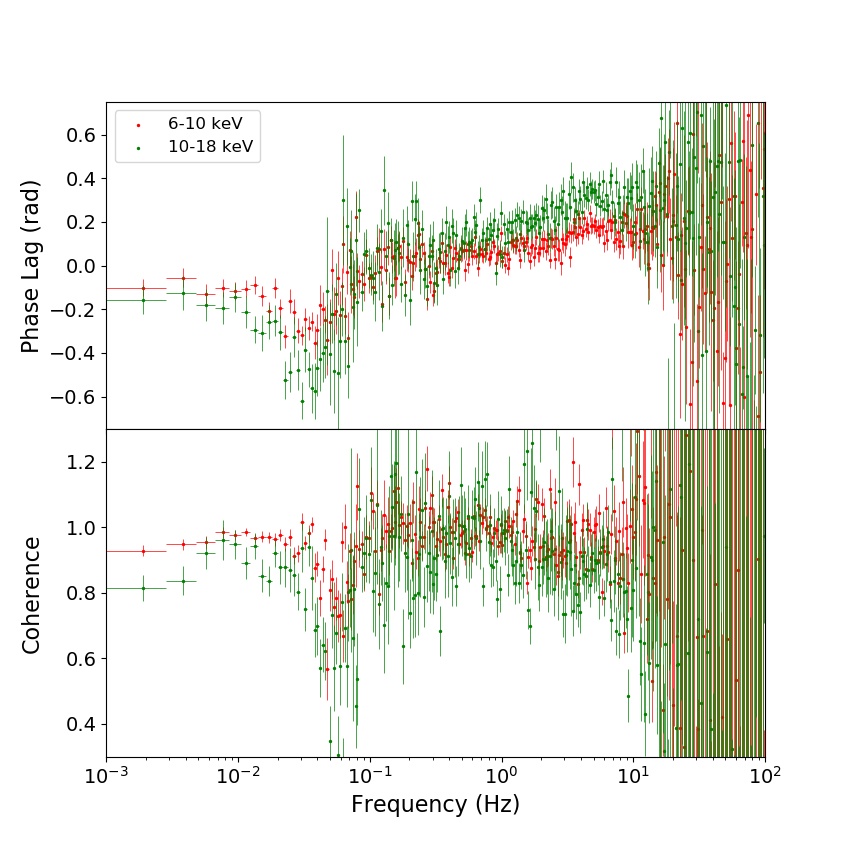}
    \caption{The phase-lag frequency spectra and coherence function for observation 1210 of Cyg X-1, showing the narrow dip in the coherence at $\sim0.05$ Hz, for two different subject bands, $6-10$ keV and $10-18$ keV, taking the $3-5$ keV band as the reference band.}
    \label{fig:lcene}
\end{figure}

We further study the energy dependence of the properties of the dip for the observation 1210. Figure \ref{fig:lcene} shows the phase lags and coherence for two different subject bands of $6-10$ keV and $10-18$ keV, with the same reference band of $3-5$ keV. We find that the rms of the Lorentzian that accounts for the drop increases from $0.4\pm0.2~\%$ to $0.6\pm0.2~\%$ and $1.1\pm0.1~\%$ to $1.3\pm0.1~\%$ in the reference and subject bands, respectively.

\section{Discussion} \label{sec:discussion}

We report the discovery of a narrow drop in the $3-5$ keV vs. $6-40$ keV coherence function at $\sim 0.05$ Hz in Cyg X-1 using AstroSat/LAXPC data. Previous studies had detected similar coherence drops in the $1-6$ Hz frequency range using NICER observations, only when data below $\sim2$ keV were used. Our work is the first to identify this coherence drop at energies greater than 3 keV. By simultaneously fitting the PS in the $3-5$ keV band and $6-40$ keV band ($0.3-100$ Hz), the real and imaginary parts of the CS, and the coherence ($0.002-0.3$ Hz), we find that the dip corresponds to a narrow variability component, akin to a QPO, whose rms increases with energy. This drop of the coherence function occurs only in the hard intermediate (HI) state, which coincides with strong broadband jet activity. 
  
The power spectrum (PS) of Cyg X-1 does not show strong and narrow QPOs, but consists of a small number of broad features over the entire frequency range. These features are fitted by broad Lorentzians and have historically not been classified as QPOs as they have a quality factor $Q=\nu_0/\Delta~<~2$. The number of Lorentzians required to fit the PS varies from two to four depending on the energy and frequency ranges afforded by the data or chosen by the author \citep[e.g.,][]{nowak00, pottschmidt03, axelsson05,kushwaha21,konig24}. In addition, there may be variability components that are only significantly detected in the CS and not in the PS \citep[][]{fogantini25}. From our analysis, we find that the PS of Cyg X-1 consists of five main components in the $10^{-3}-10^2$ Hz range, with different components dominating the total power in different states (Figure \ref{fig:rmsevo}). All of these five components were detected by \citet{pottschmidt03} in the hard state with RXTE/PCA, although they did not recognize the lowest-frequency Lorentzian as a separate component and could fit a power law equally well. While this component is indeed very weak in the hard state (observation 0890), its amplitude increases in the intermediate states and almost dominates the total power in the soft states (Figure \ref{fig:rmsevo} and \ref{fig:pclc1592}). 

The PS for observations in the soft state are usually fitted by a cutoff power law, sometimes along with a Lorentzian at higher frequencies \citep[e.g.,][]{pottschmidt03,axelsson05}. However, the coherence functions in some soft state observations, like 1540 and 1592, have a small drop around 0.01 Hz. This requires, if the assumptions of \citet{mendez24} are correct, two overlapping Lorentzians (1 and 2), as only one (coherent) component (e.g., a power law) would lead to unity coherence over a broad frequency range. Furthermore, the overall coherence level in 1592 is less than 1, indicating interference of multiple components across all frequencies. We demonstrate this by fitting the soft state observation 1592 with a model consisting of a cutoff power law and a Lorentzian with our frequency-segmented fitting scheme (see Appendix \ref{sec:compare1592} for more details).

In four observations (1094, 1122, 1180, and 1210), we detect between one and three additional components. Although these components are not always significant in the PS of both the energy bands or in the CS (Section \ref{sec:addlor}), they are necessary to adequately predict the phase lag and the coherence function. We note that several additional Lorentzians were also required by \citet{pottschmidt03} to improve the fit statistics, without looking at the CS, the lag spectrum, and the coherence function. An interesting result of our work is the discovery of the dip in the coherence at $\sim0.05$ Hz when using the hard X-ray bands (3-40 keV). This dip is accompanied by a drop in the phase lags to large negative values at the same frequency. This feature is most prominently detected in 1210 (Figures \ref{fig:lagcoh} and \ref{fig:pclc1210}), but is also significant in 1180 (Figure \ref{fig:lagcoh}) at a slightly lower frequency. The Lorentzians that fit the dip in both the observations have a low rms amplitude of $\lesssim1~\%$, which increases with energy (Figure \ref{fig:lcene}). 

Being a scarcely studied statistic, there are not many works that focus on the systematic evolution of coherence and its physical origin. Drops similar to the one we have detected here were only reported once with RXTE/PCA (similar energy bands used in this work) in the soft state of GRS 1915$+$105 \citep{jijf03}. A similar drop in coherence can be noticed at lower frequencies (~0.01 Hz) of Cyg X-1 in Nowak et al. (1999, see their Figure 5), although it was not reported. Recent studies with NICER have revived interest in the topic with the discovery of dips in the coherence in MAXI J1820$+070$ \citep{mendez24} and Cyg X-1, MAXI J1348$-630$ and AT 2019wey \citep{konig24} at $\sim 2$ Hz. Subsequently, systematic studies covering the transition from the soft to hard state in these sources uncovered many epochs with such drops in coherence \citep{bellavita25, fogantini25}. These features are essentially components that have a large imaginary part and a small real part in the CS. However, the additional component required to fit the drop in coherence in Cyg X-1 with LAXPC data is significant in the subject-band PS and has a much stronger real part than the imaginary part of the CS (Section \ref{sec:addlor}). This drop in coherence is primarily caused by a large difference in the phase lags between the component that fits the drop and other significant components at the same frequency. \citet{alabarta25} also found that in MAXI J1348--630 the dip in coherence was associated with a Type-C QPO detected in the PS.

Although the coherence dips in the aforementioned studies are morphologically similar to the ones we detected in Cyg X-1, there are many differences. Those studies were carried out with NICER, using a soft reference band of $\sim0.3-2$ keV, and showed that in Cyg X-1 the coherence does not drop when the reference band is greater than 1.6 keV \citep{konig24, fogantini25}. This prompted the authors to associate the feature with the properties of the accretion disk. As shown in Appendix \ref{sec:spec}, the contribution of disk emission in the $3-5$ keV reference band considered in this work is $<5~\%$. The coherence in the LAXPC bands reaches values as low as $\sim0.6$, a drop from unity that is about a factor of two larger than in the NICER bands (the coherence in those cases drops to $\sim0.8$). Moreover, the drop in the coherence occurs at a much lower frequency with LAXPC ($\sim 0.05$ Hz) than with NICER ($\sim1-6$ Hz). The exposure times and length of GTI of the NICER data are generally not long enough to allow us to study a similar drop at such a low frequency. 

The coherence function gives the degree of linear correlation between the variability in two energy bands \citep{bendat10, vaughan97}. While unity coherence implies a strong correlation, zero coherence indicates that the variability in the two bands are completely uncorrelated. \citet{huaxm97} studied the different avenues that could reduce the coherence in a scheme where hard photons were produced by Comptonization of soft photons. They found that the coherence can drop below unity in many ways such as increasing the seed-photon temperature, decreasing the electron temperature in the Comptonizing medium, or increasing the density and optical depth of the medium. Unlike the narrow dip of the coherence that we measure at $\sim0.05$ Hz, the drop in coherence in their model takes place over a broad frequency range. Such broadband drops are similar to the low-frequency roll-over that we detect in some observations (Figure \ref{fig:lagcoh}). The phase lags predicted from their model also decrease with frequency, opposite to what we detect in all the states (Figure \ref{fig:lagcoh}). Moreover, any perturbation in the seed photons for Comptonization is unlikely to produce the dip that we detect because our reference band ($3-5$ keV) has very little contribution from the seed photons ($kT_{in} \sim0.4$ keV).

Regardless of the origin of the variability, the coherence will be reduced if two or more components with different amplitudes and phase lags overlap in frequency \citep{vaughan97, nowak99a}. The reduction in coherence at frequencies below $\sim0.1$ Hz could result from an overlapping of two or more uncorrelated variability components from a stratified Comptonizing medium. Since our fitting is also based on the same assumptions as in \citet{nowak99a}, we find that the narrow drop in coherence at 0.05 Hz requires the presence of a narrow Lorentzian at that frequency. The large amplitude of this dip, in conjunction with a large drop in the phase lags, suggests that this extra component could have properties different from rest of the Comptonizing media, and perhaps a distinct origin. That component should also be transient, appearing only in certain phases of the activity. Therefore, by studying the properties of that specific phase of the outburst when the dip occurs, we can get an idea about its origin. 

Using 15 years of regular monitoring data with INTEGRAL,  \citet{lubinski20} divided the states in Cyg X-1 into six different accretion modes. This appears as six specific clusters in the photon index (in $30-100$ keV) - flux (in $22-100$ keV) plane \citep[see Figure 3 of][]{lubinski20}. There is an overlap in the photon index between the hard intermediate (HI) and soft intermediate (SI) states, but the two states are separated by a flux level of $75\times 10^{-10}$ erg s$^{-1}$ cm$^{-2}$. In addition to differences in flux and photon index, the states also have different variability amplitudes and different levels of radio flux \citep{lubinski20}. While all the three hard states have higher levels of radio flux than the three soft states, the peak radio flux and radio variability increases from pure hard (PH) to transitional hard (TH) and is the highest in the HI state. Not only is the radio variability high, the polarization degree in the $100-380$ keV band measured with CZTI is also the highest in the HI state \citep[$23\pm4\%$ at $>5\sigma$ confidence;][]{chattopadhyay24}. This high polarization degree was suggested to be consistent with the synchrotron emission from jets. Figure \ref{fig:czti} shows the position of our observations in the photon index - flux plane overlaid on Figure 1 of \citet{chattopadhyay24}. 

The two observations (1180 and 1210), where we detect the dip in coherence, lie in the HI state, concurrent with significant polarization measurements and high radio variability. Not only that, the four observations (1094, 1122, 1180, and 1210), which require the additional Lorentzians - beyond the five main components - also occur in the HI state. It is important to note here that the change in coherence pattern and their association with spectral states is not restricted to hidden QPOs \citep[e.g.,][]{fogantini25}. \citet{alabarta25} found that in MAXI J1348$-630$ there is an increase in radio emission during the epoch when the Type-C QPOs resulted in a dip in the coherence. Even the non-linear variability patterns at the QPO frequencies, explored through bicoherence studies, have been found to be correlated with spectral states and jet properties in BH LMXBs \citep[see e.g.,][and references therein]{arur22}. These findings suggest that, apart from the main Comptonizing medium, the base of the jet could also act as an additional medium for Comptonization. Such hybrid geometries for the corona have been suggested by some recent spectral-timing studies \citep[e.g.,][]{garciaf21, mendez22, rout23b}. 

Since the dip in coherence is associated with a narrow Lorentzian like a QPO, its origin is perhaps similar to that of a QPO. The existence of this hidden QPO, along with other broad components, indicates that the PS consists of individual additive components. If, on the contrary, the PS consisted of only one broadband variability component \citep[e.g.,][]{kawamura22}, it would not be possible to explain the QPO-like features detected through the coherence dips. This is strengthened by the fact that there are two such components, one at $\sim0.05$ Hz detected with LAXPC and another at $\sim2$ Hz detected with NICER, which are likely contemporaneous with each other as they occur at the same position in the HID.

In summary, we studied the power and cross spectral properties of Cyg X-1 in different accretion modes using data from AstroSat/LAXPC. The most important result of this work is the discovery of a narrow drop in coherence at $\sim0.05$ Hz. Using a novel technique of simultaneous frequency-segmented fitting of parts of the power spectra, cross spectra, and the coherence function with a multi-Lorentzian model, we find that the dip in the coherence is produced by a QPO. This QPO occurs in the hard intermediate state \citep[as defined by][]{lubinski20}, which shows high radio variability and a significantly high degree of linear polarization that is consistent with jet synchrotron emission \citep{chattopadhyay24}. We suggest that the dip in coherence is caused by interference of the variability arising from Comptonization in the jet with other main components.

\begin{acknowledgments}

The authors acknowledge constructive feedback on the manuscript by the reviewer. This research is based upon work supported by Tamkeen under the NYU Abu Dhabi Research Institute grant CASS. SKR acknowledges the support of the COSPAR fellowship programme for partially funding a visit to the University of Groningen. MM acknowledge the research programme Athena with project number 184.034.002, which is (partly) financed by the Dutch Research Council (NWO). FG acknowledges support by PIBAA 1275 and PIP 0113 (CONICET). FG was also supported by grant PID2022-136828NB-C42 funded by the Spanish MCIN/AEI/ 10.13039/501100011033 and “ERDF A way of making Europe”.

\end{acknowledgments}

\facilities{AstroSat, MAXI}

\software{\textsc{xspec} \citep{arnaud96}, Matplotlib \citep{hunter07}}


\appendix

\section{Methodology} \label{sec:method}

Let $x(t)$ and $y(t)$ be correlated lightcurves in two energy bands, i.e., the reference and subject bands, with $X(\nu)$ and $Y(\nu)$ being their respective Fourier transforms. The PS in the two bands can be modeled as follows.

\begin{equation}
    \begin{gathered}
        G_{xx}(\nu) ~= ~\langle X_i^{\ast}X_i\rangle ~=~ \sum_{i=1}^{n} A_i ~ L_i(\nu) \\
	    G_{yy}(\nu) ~= ~ \langle Y_i^{\ast}Y_i\rangle ~ = ~\sum_{i=1}^{n} B_i ~ L_i(\nu),  
    \end{gathered}
    \label{eq:ps}
\end{equation}

\noindent where $L_i(\nu)$ is a Lorentzian function with two parameters, the centroid frequency ($\nu_{0,i}$) and FWHM ($\Delta_i$), and $A_i$, $B_i$ are the integrated power of each Lorentzian component in the reference and subject bands, respectively. These components should also manifest themselves in the CS. \citet{mendez24} showed that if the variability components are coherent in the two energy bands, but incoherent with one another then the CS will also be a linear combination of Lorentzians, given by 

\begin{equation}
\begin{gathered}
    \Re[G_{xy}(\nu)] = \sum_{i=1}^{n} \sqrt{A_i B_i}~ L_i(\nu)  ~ \cos[\Delta\phi_{xy,i}(\nu)] \\
    \Im[G_{xy}(\nu)] = \sum_{i=1}^{n} \sqrt{A_i B_i} ~ L_i(\nu)  ~ \sin[\Delta\phi_{xy,i}(\nu)], 
\end{gathered} 
\label{eq:cs}
\end{equation} 

where $\Delta \phi_{xy,i}(\nu) = g_i(\nu;p_j)$ are the frequency-dependent phase lags of each Lorentzian between the two energy bands. The normalizations of each component in $\Re[G_{xy}(\nu)]$ and $\Im[G_{xy}(\nu)]$ is the square root of the product of the normalizations of the PS in the two energy bands. Thus, while fitting, we link the normalizations from the two bands with the normalizations of the CS, i.e., $C_i=\sqrt{A_iB_i}$. From Equation \ref{eq:cs}, we can derive the expressions for the total phase lag and coherence as follows:

\begin{equation}
\begin{gathered}
     \Delta \phi(\nu) = \tan^{-1} \left(\frac{\sum_{i=1}^{n}\sqrt{A_iB_i}~L_i(\nu)~\sin[g_i(\nu;p_j)]}{\sum_{i=1}^{n}\sqrt{A_iB_i}~L_i(\nu)~\cos[g_i(\nu;p_j)]}\right)  
\end{gathered}
\label{eq:lag}
\end{equation}

\begin{equation}
\begin{gathered}
    \gamma^2_{xy}(\nu) = \frac{|G_{xy}(\nu)|^2}{G_{xx}(\nu)G_{yy}(\nu)} = \\
    \frac{\left(\sum_{i=1}^n\sqrt{A_iB_i} ~ L_i(\nu)~ \cos[g_i(\nu;p_j)]\right)^2+\left(\sum_{i=1}^n\sqrt{A_iB_i} ~ L_i(\nu)~ \sin[g_i(\nu;p_j)]\right)^2}{\left(\sum_{i=1}^{n} A_i ~ L_i(\nu)\right) \left(\sum_{i=1}^{n} B_i ~ L_i(\nu)\right)}
    \end{gathered}
    \label{eq:coh}
\end{equation}

To set up the simultaneous fitting procedure, we need to specify models, $g_i(\nu;p_j)$, for the phase lag frequency spectrum of each Lorentzian. The method is independent of the functional form of the phase lag model, albeit the results depend on the choice. Hence, we use the simplest case of a constant model $g_i(\nu;p_i)=2\pi k_i$ with a single parameter $p_i=k_i$. Using Equations \ref{eq:lag} and \ref{eq:coh}, we can predict the phase lag and coherence, from the same set of Lorentzians that fit the two PS (Equation \ref{eq:ps}) and CS (Equation \ref{eq:cs}). This predictive potential of the method can result in the discovery of variability components which may be significant in the CS or have a significant effect on the coherence function, but not in the PS \citep{mendez24}. 

The Levenberg-Marquardt fitting algorithm, used in \textsc{xspec}, becomes unstable when the free parameters differ by more than a few orders of magnitude. This can occur for the real and imaginary parts of the CS, especially when the phase lag is close to zero. To improve the stability of the fits, we rotated all the cross vectors by $45^\circ$ before fitting. This ensures that both the real and imaginary parts have comparable magnitudes for zero phase lags \citep[see][]{mendez24}. We include this rotation in the model definitions, so it has no impact on the fit parameters. 

\section{Dead time correction} \label{sec:deadtime}

Deadtime refers to the duration when the detector is reading out an event and is thus unable to detect new events. This mainly results in a decrease of the observed count rate compared to the actual count rate by a factor of $1/(1+\tau_d R_0)$, where $\tau_d$ and $R_0$ are, respectively, the deadtime and incident count rate \citep{zhangw95}. Deadtime introduces spurious correlations between event arrival times and distorts the power spectrum, especially introducing a frequency-dependence to the white noise. The count rates at frequencies greater than $1/\tau_d$ are systematically suppressed compared to the rate at frequencies less than $1/\tau_d$. This results in a modulation of the high frequency noise with nodes at roughly $1/\tau_d$ and its harmonics \citep{zhangw95}. Deadtime can be modeled if it remains constant with time and only source events are responsible for it \citep[e.g.,][]{zhangw95}. However, like most detectors the deadtime of LAXPC varies with the type of event \citep{yadav16b}. Therefore, removal of the true Poisson level is difficult. The traditional method involves subtracting the average power from a high frequency range, e.g., 500$-$1000 Hz, where the source is not considered to be variable. While this method works reasonably well for basic power spectral analyses, such as measuring the frequency and power of strong variability components (e.g., QPOs and red noise in hard states), it fails in accurate estimation of higher order statistics like coherence, especially at high Fourier frequency \citep[see][]{vaughan97}.    

We therefore computed the power and cross spectra by cross correlating the data of the two LAXPC units (10 and 20) to remove the effects of deadtime from all Fourier products. The cross-spectrum between the two LAXPCs is a complex quantity, where the real part (or cospectrum) represents the signal components that are in phase between the data from the two detectors, while the imaginary part (or quadrature) represents the signal components that are out of phase between the two detectors \citep{bendat10, bachetti15}. Therefore, by considering only the cospectrum we can eliminate all uncorrelated, or out-of-phase variability, including that due to deadtime. The analog of a PS is the cospectrum evaluated from LAXPC10 and LAXPC20 in the same energy bands. For the Fourier transforms $X_i(\nu)$ and $Y_i(\nu)$ of two lightcurves, with $i=1,2$ representing the two LAXPCs, the cospectrum in the two bands is given by $\Re[\langle X^\ast_1(\nu)X_2(\nu)\rangle]$ and $\Re[\langle Y^\ast_1(\nu)Y_2(\nu)\rangle]$. Since the Poisson noise and the deadtime in the two detectors are uncorrelated, they are automatically eliminated from the cospectrum. Thus, it is no longer needed to subtract the Poisson noise, which is difficult owing to the deadtime effect. Further, we evaluated the cross spectrum between the two energy bands by selecting the two bands from the two LAXPCs. The real and imaginary parts of this cross vector, i.e., the cross cospectrum and cross quadrature, are then the average of the real and imaginary parts of $\langle X^\ast_1(\nu)Y_2(\nu) \rangle$ and $\langle X^\ast_2(\nu)Y_1(\nu)\rangle$, respectively. Similarly, the cross-phase lag is the average of $\tan^{-1} \left( \frac{\Im [\langle X^\ast_1(\nu)Y_2(\nu) \rangle]}{\Re[\langle X^\ast_1(\nu)Y_2(\nu) \rangle]} \right)$ and $\tan^{-1} \left( \frac{\Im [\langle X^\ast_2(\nu)Y_1(\nu)\rangle]}{\Re[\langle X^\ast_2(\nu)Y_1(\nu)\rangle]} \right)$ and the cross-coherence is the average of $\frac{\Re[\langle X^\ast_1(\nu)Y_2(\nu) \rangle]^2~+~\Im[\langle X^\ast_1(\nu)Y_2(\nu) \rangle]^2}{\Re[\langle X^\ast_1(\nu)X_2(\nu) \rangle]~\times ~ \Re[\langle Y^\ast_1(\nu)Y_2(\nu)\rangle]}$ and $\frac{\Re[\langle X^\ast_2(\nu)Y_1(\nu)\rangle]^2~+~\Im[\langle X^\ast_2(\nu)Y_1(\nu)\rangle]^2}{\Re[\langle X^\ast_1(\nu)X_2(\nu) \rangle]~\times ~ \Re[\langle Y^\ast_1(\nu)Y_2(\nu)\rangle]}$, where the subscripts 1 and 2 represent the two LAXPCs. 
This is crucial for maintaining coherence, as deadtime introduces correlations between channels, which reduce the coherence at high Fourier frequencies. Whenever the coherence is low and the power high, the uncertainties on coherence were evaluated from Equation 9 of \citet{vaughan97}. More details on this method, the uncertainties for different levels of coherence and power and their implications will be presented in a separate paper (Garc{\'i}a et al., in prep.).

\section{Spectral fits to observation 1210} \label{sec:spec}

\begin{figure}
    \centering
    \includegraphics[scale=1]{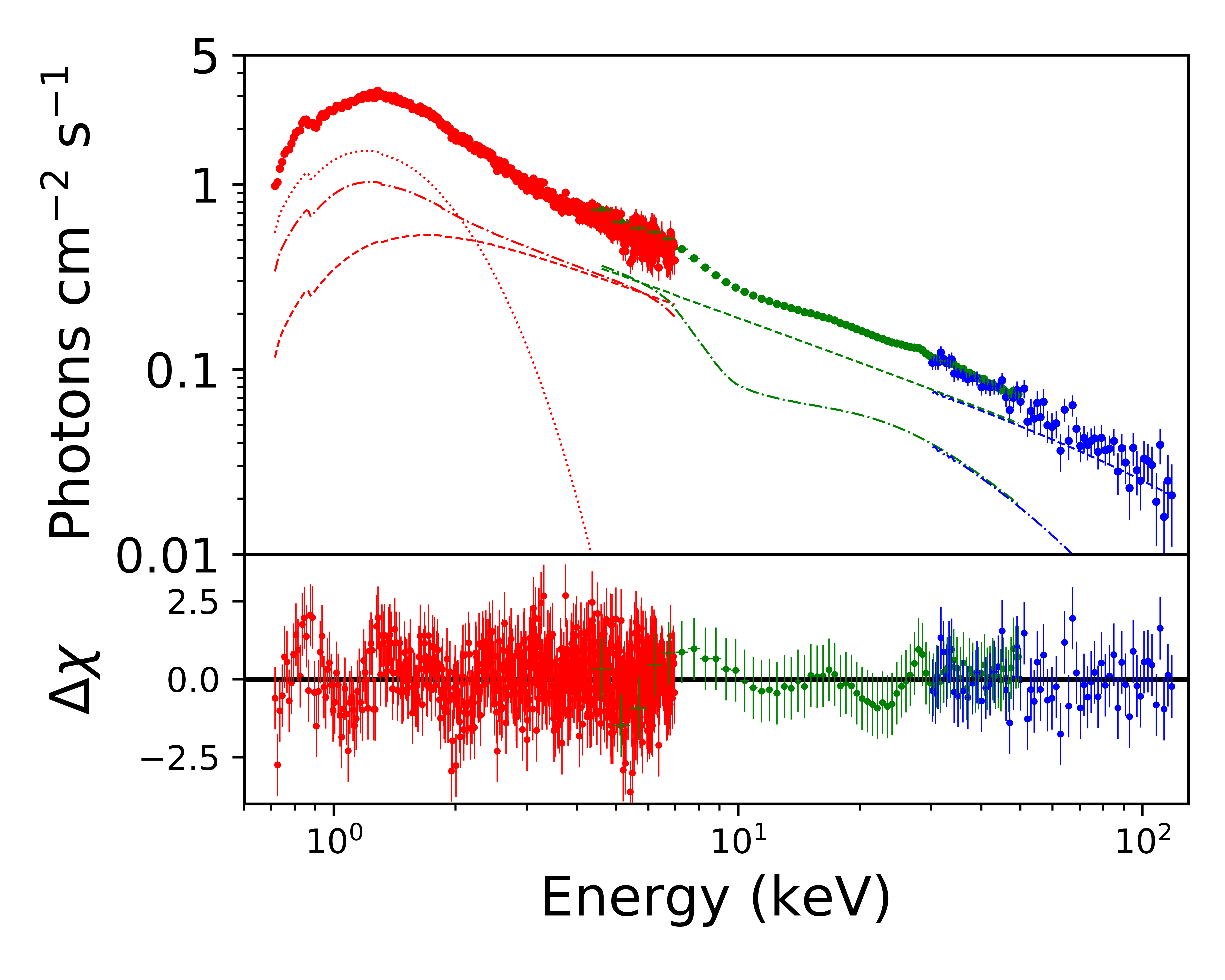}
    \caption{The joint SXT (red), LAXPC (green) and CZTI (blue) spectra and best-fitting model for observation 1210 of Cyg X-1. The individual model components are shown with dotted (disk), dashed (Comptonization) and dot-dashed (reflection) lines.}
    \label{fig:specfit}
\end{figure}

We performed broadband spectroscopy of Cyg X-1 in the $0.7-120$ keV band using all the three instruments onboard AstroSat. The reduction of LAXPC spectra was also done with the dedicated modules of the Format A pipeline. The source and background spectra were extracted from the event file using \texttt{laxpc\_make\_spectra} and \texttt{laxpc\_make\_backspectra}. The former module also generates the response files for the source and background spectra. The source spectrum was grouped to have a minimum 30 counts per channel and a systematic error of $3\%$ was added. The analysis of Soft X-ray Telescope \citep[SXT;][]{singh16} data was done using the \texttt{AS1SXTLevel2-1.4b} \footnote{\url{https://www.tifr.res.in/astrosat_sxt/sxtpipeline.html}} pipeline. Firstly, the orbit-wise event lists were merged with the \texttt{SXTMerger} \footnote{\url{https://github.com/gulabd/SXTMerger.jl}} tool. Following this, the spectrum was extracted from suitable annular regions using \textsc{xselect}. The pre-computed response and background files provided in the pipeline package were used for fitting. Since the SXT spectrum was found to be piled up, an annular extraction region with inner and outer radii of $3'$ and $15'$, respectively, was used \citep[see e.g.,][]{rout21b}. The obtained spectra were grouped to have a minimum of 30 counts per channel. 

The evolution of spectral properties of Cyg X-1 has been studied often in the past \citep[see][for data from the same epochs as ours]{kushwaha21}. In this work, we only analyzed the energy spectrum of observation 1210 to obtain the contribution of the different components to the total flux in the reference and subject bands. We fitted the joint SXT, LAXPC20 and CZTI spectra with a model consisting of an accretion disk \citep[\texttt{diskbb;}][]{makishima86}, a Comptonization \citep[\texttt{nthComp;}][]{zycki99} and a reflection component \citep[\texttt{relxillCp;}][]{dauser14}, all affected by interstellar absorption \texttt{TBabs}. For the interstellar absorption, \texttt{TBabs}, we adopted the solar abundance tables of \citet{wilms00} and the cross-section tables of \citet{verner96}. In \textsc{xspec} parlance, the model can be written as \texttt{constant * TBabs * (diskbb + nthComp + relxillCp)}. The \texttt{constant} component takes care of the cross-calibration uncertainties between the three detectors. We tied the temperature of the source of seed photons for Comptonization to the inner disk temperature ($kT_{\text{in}} = 0.40 \pm 0.01$ keV) and the photon index ($\Gamma = 1.81\pm0.02$) across the Comptonization and reflection components. The best-fitting model indicates high values for the ionization parameter ($\log \xi = 3.59^{+0.03}_{-0.07}$), disk density ($\log \text{N}>19.9$) and iron abundance ($A_{\text{Fe}} > 8.6$). This model provides a very good fit to the data. The spectra and the corresponding models are represented in Figure \ref{fig:specfit}. In the $3-5$ keV band, the disk contributes $<5~\%$ to the total flux while the Comptonization and reflection components contribute about $\sim47~\%$ and $\sim49~\%$, respectively. In the $6-40$ keV band, the Comptonization clearly dominates with about $70~\%$ of the total flux, and the reflection contributes the rest.    

\section{Power law fits to observation 1592} \label{sec:compare1592}

\begin{figure}
    \centering
    \includegraphics[scale=.6]{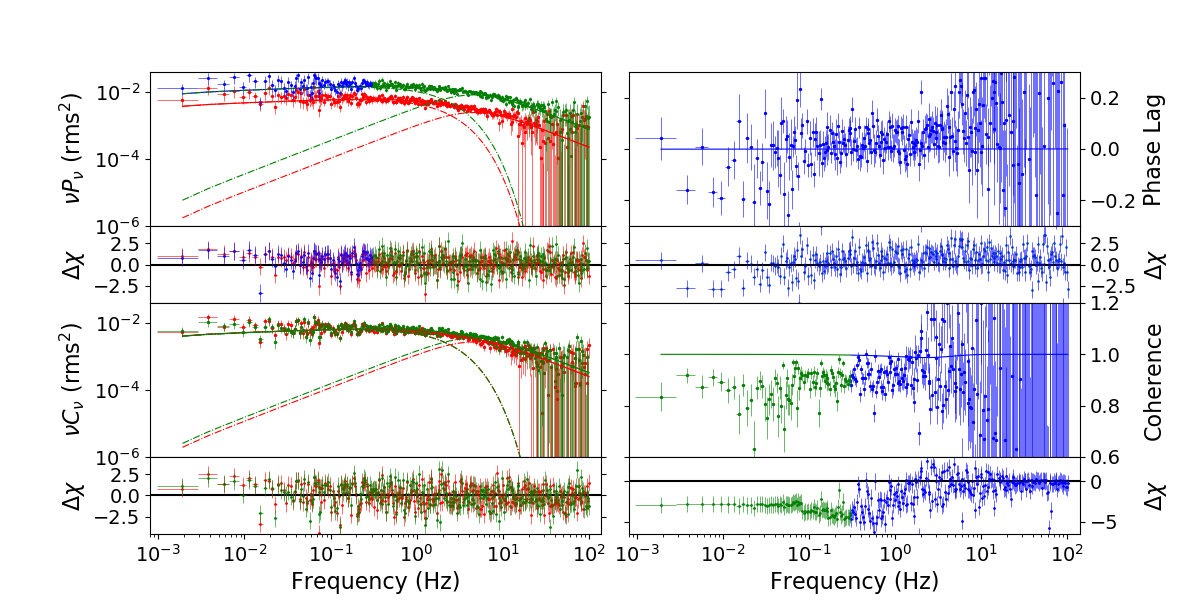}
    \caption{Same as Figure \ref{fig:pclc1210} for observation 1592, but with a model consisting of a cutoff power law and a lorentzian.}
    \label{fig:pclc1592_cpl}
\end{figure}

\begin{figure}
    \centering
    \includegraphics[scale=.6]{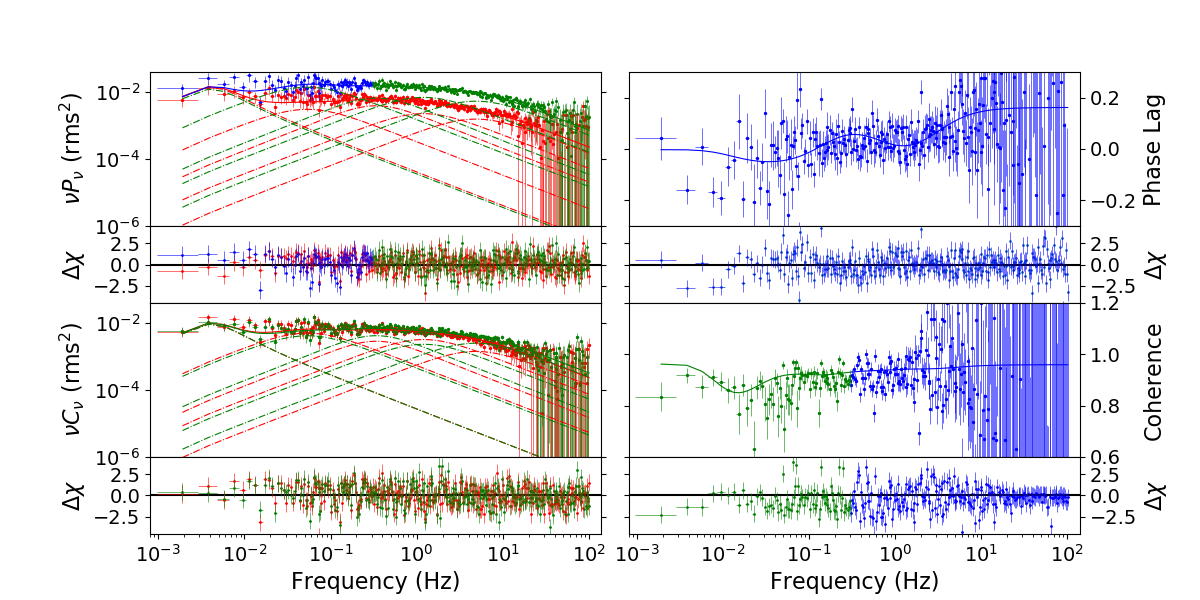}
    \caption{Same as Figure \ref{fig:pclc1210} for observation 1592.}
    \label{fig:pclc1592}
\end{figure}

Here we demonstrate that a model consisting of a power law, that is traditionally fitted to the soft state PS, cannot account for the coherence function. We use the same method as described in Appendix \ref{sec:method}, which works, in principle, for any additive component and not just a Lorentzian. We fitted observation 1592, when Cyg X-1 was in the softest state, with a model consisiting only of two components - \texttt{cutoffpl} $+$ \texttt{lorentz} - and the fitting was done with the same frequency-segmented framework described in Section \ref{sec:analysis}. The data and the best-fitting model are shown in Figure \ref{fig:pclc1592_cpl}. Although the fits to the PS and CS are reasonably good, the coherence function is neither well fitted in the $0.002-0.3$ Hz range nor well predicted in the $0.3-100$ Hz range. This is expected as the entire low frequency range of $0.002-1$ Hz is dominated by a single component, i.e., the \texttt{cutoffpl}, which under our assumption is completely coherent in the two energy bands, giving $\gamma^2=1$. In contrast, a fit with five Lorentzians not only provides a better fit to the PS and CS, but also accounts for the phase-lags and coherence function (Figure \ref{fig:pclc1592}).

The model coherence can be reduced if we relax the assumption that the variability components are coherent in the two energy bands. We did this experiment by untying the normalization of the CS from the normalizations of the two PS for the \texttt{cutoffpl} model. Despite some systematic residuals at low frequencies, the fits improved significantly as the model coherence dropped to $\sim 0.9$ in the frequency range dominated by the power law. This suggests that if the power in the soft states originates from a single component, then the process corresponding to that component is not fully coherent in the two energy bands. If Comptonization is responsible for the hard X-ray emission, then the broadband drop in the coherence could be caused by inhomogeneity in the medium, such as changes in the electron temperature or optical depth \citep[e.g.,][]{huaxm97}. However, as noted in Section \ref{sec:discussion}, these factors do not explain the sharp drop in coherence at $\sim 0.05$ Hz in the HI state.  

\newpage
\vspace*{50px}
\bibliography{sample631}{}

\begin{thebibliography}{}
\expandafter\ifx\csname natexlab\endcsname\relax\def\natexlab#1{#1}\fi
\providecommand{\url}[1]{\href{#1}{#1}}
\providecommand{\dodoi}[1]{doi:~\href{http://doi.org/#1}{\nolinkurl{#1}}}
\providecommand{\doeprint}[1]{\href{http://ascl.net/#1}{\nolinkurl{http://ascl.net/#1}}}
\providecommand{\doarXiv}[1]{\href{https://arxiv.org/abs/#1}{\nolinkurl{https://arxiv.org/abs/#1}}}

\bibitem[{{Alabarta} {et~al.}(2025){Alabarta}, {M{\'e}ndez}, {Garc{\'\i}a}, {Altamirano}, {Zhang}, {Zhang}, {Russell}, \& {K{\"o}nig}}]{alabarta25}
{Alabarta}, K., {M{\'e}ndez}, M., {Garc{\'\i}a}, F., {et~al.} 2025, \apj, 980, 251, \dodoi{10.3847/1538-4357/ada7f9}

\bibitem[{{Ar{\'e}valo} \& {Uttley}(2006)}]{arevalo06}
{Ar{\'e}valo}, P., \& {Uttley}, P. 2006, \mnras, 367, 801, \dodoi{10.1111/j.1365-2966.2006.09989.x}

\bibitem[{{Arnaud}(1996)}]{arnaud96}
{Arnaud}, K.~A. 1996, in Astronomical Society of the Pacific Conference Series, Vol. 101, Astronomical Data Analysis Software and Systems V, ed. G.~H. {Jacoby} \& J.~{Barnes}, 17

\bibitem[{{Arur} \& {Maccarone}(2022)}]{arur22}
{Arur}, K., \& {Maccarone}, T.~J. 2022, \mnras, 514, 1720, \dodoi{10.1093/mnras/stac1463}

\bibitem[{{Axelsson} {et~al.}(2005){Axelsson}, {Borgonovo}, \& {Larsson}}]{axelsson05}
{Axelsson}, M., {Borgonovo}, L., \& {Larsson}, S. 2005, \aap, 438, 999, \dodoi{10.1051/0004-6361:20042362}

\bibitem[{{Bachetti} {et~al.}(2015){Bachetti}, {Harrison}, {Cook}, {Tomsick}, {Schmid}, {Grefenstette}, {Barret}, {Boggs}, {Christensen}, {Craig}, {Fabian}, {F{\"u}rst}, {Gandhi}, {Hailey}, {Kara}, {Maccarone}, {Miller}, {Pottschmidt}, {Stern}, {Uttley}, {Walton}, {Wilms}, \& {Zhang}}]{bachetti15}
{Bachetti}, M., {Harrison}, F.~A., {Cook}, R., {et~al.} 2015, \apj, 800, 109, \dodoi{10.1088/0004-637X/800/2/109}

\bibitem[{{Bellavita} {et~al.}(2025){Bellavita}, {M{\'e}ndez}, {Garc{\'\i}a}, {Ma}, \& {K{\"o}nig}}]{bellavita25}
{Bellavita}, C., {M{\'e}ndez}, M., {Garc{\'\i}a}, F., {Ma}, R., \& {K{\"o}nig}, O. 2025, arXiv e-prints, arXiv:2502.12283, \dodoi{10.48550/arXiv.2502.12283}

\bibitem[{{Belloni} {et~al.}(2005){Belloni}, {Homan}, {Casella}, {van der Klis}, {Nespoli}, {Lewin}, {Miller}, \& {M{\'e}ndez}}]{belloni05}
{Belloni}, T., {Homan}, J., {Casella}, P., {et~al.} 2005, \aap, 440, 207, \dodoi{10.1051/0004-6361:20042457}

\bibitem[{{Belloni} {et~al.}(2002){Belloni}, {Psaltis}, \& {van der Klis}}]{belloni02}
{Belloni}, T., {Psaltis}, D., \& {van der Klis}, M. 2002, \apj, 572, 392, \dodoi{10.1086/340290}

\bibitem[{{Belloni}(2010)}]{belloni10}
{Belloni}, T.~M. 2010, {States and Transitions in Black Hole Binaries}, ed. T.~{Belloni}, Vol. 794 (Springer-Verlag, Berlin), 53, \dodoi{10.1007/978-3-540-76937-8_3}

\bibitem[{{Belloni} \& {Motta}(2016)}]{belloni16}
{Belloni}, T.~M., \& {Motta}, S.~E. 2016, {Transient Black Hole Binaries}, ed. C.~{Bambi}, Vol. 440 (Springer), 61, \dodoi{10.1007/978-3-662-52859-4_2}

\bibitem[{{Bendat} \& {Piersol}(2011)}]{bendat10}
{Bendat}, J.~S., \& {Piersol}, A.~G. 2011, {Random Data: Analysis and Measurement Procedures} ({Wiley Series in Probability and Statistics (Wiley)})

\bibitem[{{Casella} {et~al.}(2004){Casella}, {Belloni}, {Homan}, \& {Stella}}]{casella04}
{Casella}, P., {Belloni}, T., {Homan}, J., \& {Stella}, L. 2004, \aap, 426, 587, \dodoi{10.1051/0004-6361:20041231}

\bibitem[{{Chattopadhyay} {et~al.}(2024){Chattopadhyay}, {Kumar}, {Rao}, {Bhargava}, {Vadawale}, {Ratheesh}, {Dewangan}, {Bhattacharya}, {Mithun}, \& {Bhalerao}}]{chattopadhyay24}
{Chattopadhyay}, T., {Kumar}, A., {Rao}, A.~R., {et~al.} 2024, \apjl, 960, L2, \dodoi{10.3847/2041-8213/ad118d}

\bibitem[{{Cui} {et~al.}(1997){Cui}, {Zhang}, {Focke}, \& {Swank}}]{cui97b}
{Cui}, W., {Zhang}, S.~N., {Focke}, W., \& {Swank}, J.~H. 1997, \apj, 484, 383, \dodoi{10.1086/304341}

\bibitem[{{Dauser} {et~al.}(2014){Dauser}, {Garcia}, {Parker}, {Fabian}, \& {Wilms}}]{dauser14}
{Dauser}, T., {Garcia}, J., {Parker}, M.~L., {Fabian}, A.~C., \& {Wilms}, J. 2014, \mnras, 444, L100, \dodoi{10.1093/mnrasl/slu125}

\bibitem[{{Dexter} \& {Begelman}(2024)}]{dexter24}
{Dexter}, J., \& {Begelman}, M.~C. 2024, \mnras, 528, L157, \dodoi{10.1093/mnrasl/slad182}

\bibitem[{{Done} {et~al.}(2007){Done}, {Gierli{\'n}ski}, \& {Kubota}}]{done07}
{Done}, C., {Gierli{\'n}ski}, M., \& {Kubota}, A. 2007, \aapr, 15, 1, \dodoi{10.1007/s00159-007-0006-1}

\bibitem[{{Dove} {et~al.}(1997){Dove}, {Wilms}, {Maisack}, \& {Begelman}}]{dove97}
{Dove}, J.~B., {Wilms}, J., {Maisack}, M., \& {Begelman}, M.~C. 1997, \apj, 487, 759, \dodoi{10.1086/304647}

\bibitem[{{Dove} {et~al.}(1998){Dove}, {Wilms}, {Nowak}, {Vaughan}, \& {Begelman}}]{dove98}
{Dove}, J.~B., {Wilms}, J., {Nowak}, M.~A., {Vaughan}, B.~A., \& {Begelman}, M.~C. 1998, \mnras, 298, 729, \dodoi{10.1046/j.1365-8711.1998.01673.x}

\bibitem[{{Fogantini} {et~al.}(2025){Fogantini}, {Garc{\'\i}a}, {M{\'e}ndez}, {K{\"o}nig}, \& {Wilms}}]{fogantini25}
{Fogantini}, F.~A., {Garc{\'\i}a}, F., {M{\'e}ndez}, M., {K{\"o}nig}, O., \& {Wilms}, J. 2025, arXiv e-prints, arXiv:2503.03078.
\newblock \doarXiv{2503.03078}

\bibitem[{{Garc{\'\i}a} {et~al.}(2021){Garc{\'\i}a}, {M{\'e}ndez}, {Karpouzas}, {Belloni}, {Zhang}, \& {Altamirano}}]{garciaf21}
{Garc{\'\i}a}, F., {M{\'e}ndez}, M., {Karpouzas}, K., {et~al.} 2021, \mnras, 501, 3173, \dodoi{10.1093/mnras/staa3944}

\bibitem[{{Grinberg} {et~al.}(2014){Grinberg}, {Pottschmidt}, {B{\"o}ck}, {Schmid}, {Nowak}, {Uttley}, {Tomsick}, {Rodriguez}, {Hell}, {Markowitz}, {Bodaghee}, {Cadolle Bel}, {Rothschild}, \& {Wilms}}]{grinberg14}
{Grinberg}, V., {Pottschmidt}, K., {B{\"o}ck}, M., {et~al.} 2014, \aap, 565, A1, \dodoi{10.1051/0004-6361/201322969}

\bibitem[{{Homan} {et~al.}(2001){Homan}, {Wijnands}, {van der Klis}, {Belloni}, {van Paradijs}, {Klein-Wolt}, {Fender}, \& {M{\'e}ndez}}]{homan01}
{Homan}, J., {Wijnands}, R., {van der Klis}, M., {et~al.} 2001, \apjs, 132, 377, \dodoi{10.1086/318954}

\bibitem[{{Hua} {et~al.}(1997){Hua}, {Kazanas}, \& {Titarchuk}}]{huaxm97}
{Hua}, X.-M., {Kazanas}, D., \& {Titarchuk}, L. 1997, \apjl, 482, L57, \dodoi{10.1086/310695}

\bibitem[{Hunter(2007)}]{hunter07}
Hunter, J.~D. 2007, Computing in Science \& Engineering, 9, 90, \dodoi{10.1109/MCSE.2007.55}

\bibitem[{{Ingram} \& {Done}(2011)}]{ingram11}
{Ingram}, A., \& {Done}, C. 2011, \mnras, 415, 2323, \dodoi{10.1111/j.1365-2966.2011.18860.x}

\bibitem[{{Ingram} \& {Motta}(2019)}]{ingram19b}
{Ingram}, A.~R., \& {Motta}, S.~E. 2019, \nar, 85, 101524, \dodoi{10.1016/j.newar.2020.101524}

\bibitem[{{Ji} {et~al.}(2003){Ji}, {Zhang}, {Qu}, \& {Li}}]{jijf03}
{Ji}, J.~F., {Zhang}, S.~N., {Qu}, J.~L., \& {Li}, T.~P. 2003, \apjl, 584, L23, \dodoi{10.1086/368269}

\bibitem[{{Jourdain} {et~al.}(2012){Jourdain}, {Roques}, {Chauvin}, \& {Clark}}]{jourdain12}
{Jourdain}, E., {Roques}, J.~P., {Chauvin}, M., \& {Clark}, D.~J. 2012, \apj, 761, 27, \dodoi{10.1088/0004-637X/761/1/27}

\bibitem[{{Kalemci} {et~al.}(2022){Kalemci}, {Kara}, \& {Tomsick}}]{kalemci22}
{Kalemci}, E., {Kara}, E., \& {Tomsick}, J.~A. 2022, in Handbook of X-ray and Gamma-ray Astrophysics (Springer Nature Reference), 9, \dodoi{10.1007/978-981-16-4544-0_100-1}

\bibitem[{{Kara} {et~al.}(2019){Kara}, {Steiner}, {Fabian}, {Cackett}, {Uttley}, {Remillard}, {Gendreau}, {Arzoumanian}, {Altamirano}, {Eikenberry}, {Enoto}, {Homan}, {Neilsen}, \& {Stevens}}]{kara19}
{Kara}, E., {Steiner}, J.~F., {Fabian}, A.~C., {et~al.} 2019, \nat, 565, 198, \dodoi{10.1038/s41586-018-0803-x}

\bibitem[{{Karpouzas} {et~al.}(2020){Karpouzas}, {M{\'e}ndez}, {Ribeiro}, {Altamirano}, {Blaes}, \& {Garc{\'\i}a}}]{karpouzas20}
{Karpouzas}, K., {M{\'e}ndez}, M., {Ribeiro}, E.~M., {et~al.} 2020, \mnras, 492, 1399, \dodoi{10.1093/mnras/stz3502}

\bibitem[{{Kawamura} {et~al.}(2022){Kawamura}, {Axelsson}, {Done}, \& {Takahashi}}]{kawamura22}
{Kawamura}, T., {Axelsson}, M., {Done}, C., \& {Takahashi}, T. 2022, \mnras, 511, 536, \dodoi{10.1093/mnras/stac045}

\bibitem[{{K{\"o}nig} {et~al.}(2024){K{\"o}nig}, {Mastroserio}, {Dauser}, {M{\'e}ndez}, {Wang}, {Garc{\'\i}a}, {Steiner}, {Pottschmidt}, {Ballhausen}, {Connors}, {Garc{\'\i}a}, {Grinberg}, {Horn}, {Ingram}, {Kara}, {Kallman}, {Lucchini}, {Nathan}, {Nowak}, {Thalhammer}, {van der Klis}, \& {Wilms}}]{konig24}
{K{\"o}nig}, O., {Mastroserio}, G., {Dauser}, T., {et~al.} 2024, \aap, 687, A284, \dodoi{10.1051/0004-6361/202449333}

\bibitem[{{Kotze} \& {Charles}(2012)}]{kotze12}
{Kotze}, M.~M., \& {Charles}, P.~A. 2012, \mnras, 420, 1575, \dodoi{10.1111/j.1365-2966.2011.20146.x}

\bibitem[{{Krawczynski} {et~al.}(2022){Krawczynski}, {Muleri}, {Dov{\v{c}}iak}, {Veledina}, {Rodriguez Cavero}, {Svoboda}, {Ingram}, {Matt}, {Garcia}, {Loktev}, {Negro}, {Poutanen}, {Kitaguchi}, {Podgorn{\'y}}, {Rankin}, {Zhang}, {Berdyugin}, {Berdyugina}, {Bianchi}, {Blinov}, {Capitanio}, {Di Lalla}, {Draghis}, {Fabiani}, {Kagitani}, {Kravtsov}, {Kiehlmann}, {Latronico}, {Lutovinov}, {Mandarakas}, {Marin}, {Marinucci}, {Miller}, {Mizuno}, {Molkov}, {Omodei}, {Petrucci}, {Ratheesh}, {Sakanoi}, {Semena}, {Skalidis}, {Soffitta}, {Tennant}, {Thalhammer}, {Tombesi}, {Weisskopf}, {Wilms}, {Zhang}, {Agudo}, {Antonelli}, {Bachetti}, {Baldini}, {Baumgartner}, {Bellazzini}, {Bongiorno}, {Bonino}, {Brez}, {Bucciantini}, {Castellano}, {Cavazzuti}, {Ciprini}, {Costa}, {De Rosa}, {Del Monte}, {Di Gesu}, {Di Marco}, {Donnarumma}, {Doroshenko}, {Ehlert}, {Enoto}, {Evangelista}, {Ferrazzoli}, {Gunji}, {Hayashida}, {Heyl}, {Iwakiri}, {Jorstad}, {Karas}, {Kolodziejczak}, {La Monaca}, {Liodakis}, {Maldera}, {Manfreda},
  {Marscher}, {Marshall}, {Mitsuishi}, {Ng}, {O{\textquoteright}Dell}, {Oppedisano}, {Papitto}, {Pavlov}, {Peirson}, {Perri}, {Pesce-Rollins}, {Pilia}, {Possenti}, {Puccetti}, {Ramsey}, {Romani}, {Sgr{\`o}}, {Slane}, {Spandre}, {Tamagawa}, {Tavecchio}, {Taverna}, {Tawara}, {Thomas}, {Trois}, {Tsygankov}, {Turolla}, {Vink}, {Wu}, {Xie}, \& {Zane}}]{krawczynski22}
{Krawczynski}, H., {Muleri}, F., {Dov{\v{c}}iak}, M., {et~al.} 2022, Science, 378, 650, \dodoi{10.1126/science.add5399}

\bibitem[{{Kushwaha} {et~al.}(2021){Kushwaha}, {Agrawal}, \& {Nandi}}]{kushwaha21}
{Kushwaha}, A., {Agrawal}, V.~K., \& {Nandi}, A. 2021, \mnras, 507, 2602, \dodoi{10.1093/mnras/stab2258}

\bibitem[{{Kylafis} {et~al.}(2020){Kylafis}, {Reig}, \& {Papadakis}}]{kylafis20}
{Kylafis}, N.~D., {Reig}, P., \& {Papadakis}, I. 2020, \aap, 640, L16, \dodoi{10.1051/0004-6361/202038468}

\bibitem[{{Lubi{\'n}ski} {et~al.}(2020){Lubi{\'n}ski}, {Filothodoros}, {Zdziarski}, \& {Pooley}}]{lubinski20}
{Lubi{\'n}ski}, P., {Filothodoros}, A., {Zdziarski}, A.~A., \& {Pooley}, G. 2020, \apj, 896, 101, \dodoi{10.3847/1538-4357/ab9311}

\bibitem[{{Makishima} {et~al.}(1986){Makishima}, {Maejima}, {Mitsuda}, {Bradt}, {Remillard}, {Tuohy}, {Hoshi}, \& {Nakagawa}}]{makishima86}
{Makishima}, K., {Maejima}, Y., {Mitsuda}, K., {et~al.} 1986, \apj, 308, 635, \dodoi{10.1086/164534}

\bibitem[{{Matsuoka} {et~al.}(2009){Matsuoka}, {Kawasaki}, {Ueno}, {Tomida}, {Kohama}, {Suzuki}, {Adachi}, {Ishikawa}, {Mihara}, {Sugizaki}, {Isobe}, {Nakagawa}, {Tsunemi}, {Miyata}, {Kawai}, {Kataoka}, {Morii}, {Yoshida}, {Negoro}, {Nakajima}, {Ueda}, {Chujo}, {Yamaoka}, {Yamazaki}, {Nakahira}, {You}, {Ishiwata}, {Miyoshi}, {Eguchi}, {Hiroi}, {Katayama}, \& {Ebisawa}}]{matsuoka09}
{Matsuoka}, M., {Kawasaki}, K., {Ueno}, S., {et~al.} 2009, \pasj, 61, 999, \dodoi{10.1093/pasj/61.5.999}

\bibitem[{{McClintock} {et~al.}(2006){McClintock}, {Shafee}, {Narayan}, {Remillard}, {Davis}, \& {Li}}]{mcclintock06}
{McClintock}, J.~E., {Shafee}, R., {Narayan}, R., {et~al.} 2006, \apj, 652, 518, \dodoi{10.1086/508457}

\bibitem[{{M{\'e}ndez} \& {Belloni}(2021)}]{mendez21}
{M{\'e}ndez}, M., \& {Belloni}, T.~M. 2021, in Astrophysics and Space Science Library, Vol. 461, Timing Neutron Stars: Pulsations, Oscillations and Explosions, ed. T.~M. {Belloni}, M.~{M{\'e}ndez}, \& C.~{Zhang}, 263--331, \dodoi{10.1007/978-3-662-62110-3_6}

\bibitem[{{M{\'e}ndez} {et~al.}(2022){M{\'e}ndez}, {Karpouzas}, {Garc{\'\i}a}, {Zhang}, {Zhang}, {Belloni}, \& {Altamirano}}]{mendez22}
{M{\'e}ndez}, M., {Karpouzas}, K., {Garc{\'\i}a}, F., {et~al.} 2022, Nature Astronomy, 6, 577, \dodoi{10.1038/s41550-022-01617-y}

\bibitem[{{M{\'e}ndez} {et~al.}(2024){M{\'e}ndez}, {Peirano}, {Garc{\'\i}a}, {Belloni}, {Altamirano}, \& {Alabarta}}]{mendez24}
{M{\'e}ndez}, M., {Peirano}, V., {Garc{\'\i}a}, F., {et~al.} 2024, \mnras, 527, 9405, \dodoi{10.1093/mnras/stad3786}

\bibitem[{{Miller-Jones} {et~al.}(2021){Miller-Jones}, {Bahramian}, {Orosz}, {Mandel}, {Gou}, {Maccarone}, {Neijssel}, {Zhao}, {Zi{\'o}{\l}kowski}, {Reid}, {Uttley}, {Zheng}, {Byun}, {Dodson}, {Grinberg}, {Jung}, {Kim}, {Marcote}, {Markoff}, {Rioja}, {Rushton}, {Russell}, {Sivakoff}, {Tetarenko}, {Tudose}, \& {Wilms}}]{millerjones21}
{Miller-Jones}, J. C.~A., {Bahramian}, A., {Orosz}, J.~A., {et~al.} 2021, Science, 371, 1046, \dodoi{10.1126/science.abb3363}

\bibitem[{{Miyamoto} {et~al.}(1991){Miyamoto}, {Kimura}, {Kitamoto}, {Dotani}, \& {Ebisawa}}]{miyamoto91}
{Miyamoto}, S., {Kimura}, K., {Kitamoto}, S., {Dotani}, T., \& {Ebisawa}, K. 1991, \apj, 383, 784, \dodoi{10.1086/170837}

\bibitem[{{Nowak}(2000)}]{nowak00}
{Nowak}, M.~A. 2000, \mnras, 318, 361, \dodoi{10.1046/j.1365-8711.2000.03668.x}

\bibitem[{{Nowak} {et~al.}(1999){Nowak}, {Vaughan}, {Wilms}, {Dove}, \& {Begelman}}]{nowak99a}
{Nowak}, M.~A., {Vaughan}, B.~A., {Wilms}, J., {Dove}, J.~B., \& {Begelman}, M.~C. 1999, \apj, 510, 874, \dodoi{10.1086/306610}

\bibitem[{{Pottschmidt} {et~al.}(2003){Pottschmidt}, {Wilms}, {Nowak}, {Pooley}, {Gleissner}, {Heindl}, {Smith}, {Remillard}, \& {Staubert}}]{pottschmidt03}
{Pottschmidt}, K., {Wilms}, J., {Nowak}, M.~A., {et~al.} 2003, \aap, 407, 1039, \dodoi{10.1051/0004-6361:20030906}

\bibitem[{{Rout} {et~al.}(2021{\natexlab{a}}){Rout}, {M{\'e}ndez}, {Belloni}, \& {Vadawale}}]{rout21a}
{Rout}, S.~K., {M{\'e}ndez}, M., {Belloni}, T.~M., \& {Vadawale}, S. 2021{\natexlab{a}}, \mnras, 505, 1213, \dodoi{10.1093/mnras/stab1341}

\bibitem[{{Rout} {et~al.}(2023){Rout}, {M{\'e}ndez}, \& {Garc{\'\i}a}}]{rout23b}
{Rout}, S.~K., {M{\'e}ndez}, M., \& {Garc{\'\i}a}, F. 2023, \mnras, 525, 221, \dodoi{10.1093/mnras/stad2321}

\bibitem[{{Rout} {et~al.}(2025){Rout}, {Mu{\~n}oz-Darias}, {Homan}, {Armas Padilla}, {Russell}, {Alabarta}, \& {Saikia}}]{rout25}
{Rout}, S.~K., {Mu{\~n}oz-Darias}, T., {Homan}, J., {et~al.} 2025, \apj, 978, 12, \dodoi{10.3847/1538-4357/ad919f}

\bibitem[{{Rout} {et~al.}(2021{\natexlab{b}}){Rout}, {Vadawale}, {Aarthy}, {Ganesh}, {Joshi}, {Roy}, {Misra}, \& {Yadav}}]{rout21b}
{Rout}, S.~K., {Vadawale}, S.~V., {Aarthy}, E., {et~al.} 2021{\natexlab{b}}, Journal of Astrophysics and Astronomy, 42, 39, \dodoi{10.1007/s12036-021-09696-5}

\bibitem[{{Singh} {et~al.}(2016){Singh}, {Stewart}, {Chandra}, {Mukerjee}, {Kotak}, {Beardmore}, {Chitnis}, {Dewangan}, {Bhattacharyya}, {Mirza}, {Kamble}, {Navalkar}, {Shah}, {Vishwakarma}, \& {Koyande}}]{singh16}
{Singh}, K.~P., {Stewart}, G.~C., {Chandra}, S., {et~al.} 2016, in Society of Photo-Optical Instrumentation Engineers (SPIE) Conference Series, Vol. 9905, Space Telescopes and Instrumentation 2016: Ultraviolet to Gamma Ray, ed. J.-W.~A. {den Herder}, T.~{Takahashi}, \& M.~{Bautz}, 99051E, \dodoi{10.1117/12.2235309}

\bibitem[{{Tananbaum} {et~al.}(1972){Tananbaum}, {Gursky}, {Kellogg}, {Giacconi}, \& {Jones}}]{tananbaum72}
{Tananbaum}, H., {Gursky}, H., {Kellogg}, E., {Giacconi}, R., \& {Jones}, C. 1972, \apjl, 177, L5, \dodoi{10.1086/181042}

\bibitem[{{Vadawale} {et~al.}(2016){Vadawale}, {Rao}, {Bhattacharya}, {Bhalerao}, {Dewangan}, {Vibhute}, {Mithun}, {Chattopadhyay}, \& {Sreekumar}}]{vadawale16}
{Vadawale}, S.~V., {Rao}, A.~R., {Bhattacharya}, D., {et~al.} 2016, in Society of Photo-Optical Instrumentation Engineers (SPIE) Conference Series, Vol. 9905, Space Telescopes and Instrumentation 2016: Ultraviolet to Gamma Ray, ed. J.-W.~A. {den Herder}, T.~{Takahashi}, \& M.~{Bautz}, 99051G, \dodoi{10.1117/12.2235373}

\bibitem[{{van der Klis}(1989)}]{vanderklis89a}
{van der Klis}, M. 1989, \araa, 27, 517, \dodoi{10.1146/annurev.aa.27.090189.002505}

\bibitem[{{van der Klis}(1994)}]{vanderklis94}
---. 1994, \apjs, 92, 511, \dodoi{10.1086/192006}

\bibitem[{{Vaughan} \& {Nowak}(1997)}]{vaughan97}
{Vaughan}, B.~A., \& {Nowak}, M.~A. 1997, \apjl, 474, L43, \dodoi{10.1086/310430}

\bibitem[{{Verner} {et~al.}(1996){Verner}, {Ferland}, {Korista}, \& {Yakovlev}}]{verner96}
{Verner}, D.~A., {Ferland}, G.~J., {Korista}, K.~T., \& {Yakovlev}, D.~G. 1996, \apj, 465, 487, \dodoi{10.1086/177435}

\bibitem[{{Walborn}(1973)}]{walborn73}
{Walborn}, N.~R. 1973, \apjl, 179, L123, \dodoi{10.1086/181131}

\bibitem[{{Wilms} {et~al.}(2000){Wilms}, {Allen}, \& {McCray}}]{wilms00}
{Wilms}, J., {Allen}, A., \& {McCray}, R. 2000, \apj, 542, 914, \dodoi{10.1086/317016}

\bibitem[{{Wilms} {et~al.}(2006){Wilms}, {Nowak}, {Pottschmidt}, {Pooley}, \& {Fritz}}]{wilms06}
{Wilms}, J., {Nowak}, M.~A., {Pottschmidt}, K., {Pooley}, G.~G., \& {Fritz}, S. 2006, \aap, 447, 245, \dodoi{10.1051/0004-6361:20053938}

\bibitem[{{Yadav} {et~al.}(2016{\natexlab{a}}){Yadav}, {Misra}, {Verdhan Chauhan}, {Agrawal}, {Antia}, {Pahari}, {Dedhia}, {Katoch}, {Madhwani}, {Manchanda}, {Paul}, {Shah}, \& {Ishwara-Chandra}}]{yadav16b}
{Yadav}, J.~S., {Misra}, R., {Verdhan Chauhan}, J., {et~al.} 2016{\natexlab{a}}, \apj, 833, 27, \dodoi{10.3847/0004-637X/833/1/27}

\bibitem[{{Yadav} {et~al.}(2016{\natexlab{b}}){Yadav}, {Agrawal}, {Antia}, {Chauhan}, {Dedhia}, {Katoch}, {Madhwani}, {Manchanda}, {Misra}, {Pahari}, {Paul}, \& {Shah}}]{yadav16a}
{Yadav}, J.~S., {Agrawal}, P.~C., {Antia}, H.~M., {et~al.} 2016{\natexlab{b}}, in Society of Photo-Optical Instrumentation Engineers (SPIE) Conference Series, Vol. 9905, Space Telescopes and Instrumentation 2016: Ultraviolet to Gamma Ray, ed. J.-W.~A. {den Herder}, T.~{Takahashi}, \& M.~{Bautz}, 99051D, \dodoi{10.1117/12.2231857}

\bibitem[{{Zhang} {et~al.}(1995){Zhang}, {Jahoda}, {Swank}, {Morgan}, \& {Giles}}]{zhangw95}
{Zhang}, W., {Jahoda}, K., {Swank}, J.~H., {Morgan}, E.~H., \& {Giles}, A.~B. 1995, \apj, 449, 930, \dodoi{10.1086/176111}

\bibitem[{{Zhou} {et~al.}(2022){Zhou}, {Grinberg}, {Bu}, {Santangelo}, {Cangemi}, {Diez}, {K{\"o}nig}, {Ji}, {Nowak}, {Pottschmidt}, {Rodriguez}, {Wilms}, {Zhang}, {Qu}, \& {Zhang}}]{zhoum22}
{Zhou}, M., {Grinberg}, V., {Bu}, Q.~C., {et~al.} 2022, \aap, 666, A172, \dodoi{10.1051/0004-6361/202244240}

\bibitem[{{{\.Z}ycki} {et~al.}(1999){{\.Z}ycki}, {Done}, \& {Smith}}]{zycki99}
{{\.Z}ycki}, P.~T., {Done}, C., \& {Smith}, D.~A. 1999, \mnras, 309, 561, \dodoi{10.1046/j.1365-8711.1999.02885.x}

\end{thebibliography}
\bibliographystyle{aasjournal}

\end{document}